\documentclass[reqno]{amsart}
\usepackage{amsfonts,latexsym,enumerate}
\usepackage{amsmath}
\usepackage{amscd}
\usepackage{float,amsmath,amssymb,mathrsfs,bm,multirow,graphics}
\usepackage[dvips]{graphicx}
\usepackage{overpic}
\usepackage{amsaddr}
\usepackage[numbers,sort&compress]{natbib}
\usepackage{tikz}
\usepackage{epstopdf}
\usepackage{subfigure}
\usepackage{enumerate}

\numberwithin{equation}{section}

\begin{document}
	
	\title[Modulation theory of soliton$-$mean flow in KdV equation]
	{Modulation theory of soliton$-$mean flow in KdV equation with box type initial data}
	
	\author{Ruizhi Gong, Deng-Shan Wang$^{\dagger}$}
	\address{Laboratory of Mathematics and Complex Systems (Ministry of Education),  \\ School of Mathematical Sciences, Beijing Normal University, Beijing 100875, China.}
	\email{dswang@bnu.edu.cn}

	
	\date{\today}
	
	
	
\begin{abstract}
For the KdV equation with box type initial data, the interaction between a trial soliton and large-scale dispersive mean flow is studied theoretically and numerically. The pure box initial value can cause rarefaction wave and dispersive shock wave, and can create an area of soliton train. The key to the interaction of soliton and mean flow is that the dynamic evolutions of the mean flow and the local soliton can be described by the same modulation system. The soliton modulation system is derived from the degenerations of the two-genus Whitham modulation system. Considering the influence of rarefaction wave, dispersive shock wave and soliton train on the trial soliton, in the framework of Whitham modulation theory, the equation describing the soliton trajectory and the changes in amplitude and phase shift are given explicitly. The predicted results are compared with the numerical simulations, which verifies the corrections of the theoretical analysis. The exotic interaction phenomena between soliton and mean flow found in this work have broad applications to shallow water soliton propagations and real soliton experiments in fluid dynamics.
   	
\end{abstract}
	
\maketitle


\section{Introduction}

The Korteweg-de Vries (KdV) equation
\begin{equation}
\label{KdV}
u_t+6uu_x+u_{xxx}=0
\end{equation}
is a basic mathematical model describing the propagation of weakly nonlinear long waves in dispersive media \cite{Zabusky-1965,Lax-1968}. It is particularly important as a prototypical example of an exactly solvable nonlinear system whose complete integrability can be guaranteed by inverse scattering transform \cite{Gardner-1967} and an infinite number of conservation laws \cite{Miura-1976}.
Solitons and nonlinear periodic waves of this equation are two types well-known exact solutions and have been extensively studied during the past years. For Riemann problems with step initial values, Whitham modulation theory \cite{Whitham-1965}-\cite{Wang-Stud-2022} has been widely used as one of the main research methods. In fact, the dispersive hydrodynamic theory described by the integrable nonlinear dispersion equation is significantly different from the dissipative hydrodynamic theory. Whitham modulation theory asymptotically describes this multi-scale nonlinear dynamic behavior by averaging the fast oscillations in the dispersive shock wave. In the framework of Whitham modulation theory, the slow evolution of wave amplitude and wavelength can be described by a system of first-order quasi-linear partial differential equations, which is called Whitham equation. The ``modulation" here describes the slow change of the hyperelliptic Riemann surface associated with the finite gap solution of the spectrum. The modulation solution obtained by Whitham modulation theory is locally approximated by the solution of the initial problem in the original dispersive fluid dynamics.
\par
The dispersive hydrodynamics of the KdV equation considers multi-scale nonlinear wave solutions with initial conditions. The scales here include: oscillation scale (soliton width or wavelength of periodic traveling wave) and hydrodynamic scale (the slowly varying oscillatory amplitude of dispersive shock wave). In nonlinear wave systems, the interaction between small-scale dispersive waves and large-scale mean flows is a common problem in fluid mechanics with many important applications \cite{Oliver-2009}. The initial value conditions that this paper focuses on are as follows:
\begin{equation}
\label{Initial condition}
u(x,0)=u_0=U+s_0,
\end{equation}
where
\begin{equation}
\label{trial}
s_0=s(x,0;x_0)=a\mathrm{sech}^2\left(\sqrt{\frac{a}{2}}(x-x_0)\right),
\end{equation}
and
\begin{equation}
\label{box}
U=\left\{
\begin{aligned}
&U_0>0,  &0<x<l,\\
&0,  &{\rm other ~else}.\\
\end{aligned}
\right.
\end{equation}
The square barrier (\ref{box}) in this initial condition is called the box initial data. It seems that it can be regarded as the splicing of two step-like initial values with opposite monotonicity, but in fact it is significantly different from a single discontinuous step-like initial value studied in \cite{Gurevich-1974,Luo-stud-2023}. The square barrier (\ref{box}) is an important potential in quantum mechanics \cite{Hirschfelder-1974} and related fields \cite{Jenkins-2014}. For the KdV equation (\ref{KdV}), because of this difference it is precise that the interaction between the rarefaction wave and the dispersive shock wave generated by the discontinuities at points $x=0,l$ occurs.
\par
The interaction problem of soliton-mean flow that we are concerned about is the interaction caused by the trial soliton (\ref{trial}) being shot into the initial box and subsequently evolving. It must be noted that the tunneling problem considered here is not a traditional particle tunneling barrier problem. The potential barrier is no longer stationary but satisfies the same modulation equation as the soliton. The background mean flow generated by the evolution of the potential barrier over time makes the problem interesting. We broadly understand solitons as a coherent, particle-like entity that can interact elastically with other solitons and dispersive radiation, and can also interact with nonlinear hydrodynamic states without fission or radiation, albeit with different amplitudes due to variations in background mean flow. The key to the soliton-mean flow interaction is that the dynamic evolution of the mean flow and the local solitons can be described by the same modulation system, which requires the guarantee of scale separation. In fact, the soliton-mean flow interaction is scale-separated, and the characteristic length and time scale of the soliton are much smaller than those of the mean flow. This is a natural feature of dispersive hydrodynamics.
\par
Since Maiden et al. \cite{EL-PRL-2018} discovered that dynamical mean flow and propagating local solitons can be described by the same dispersive hydrodynamic equation, there have been some new developments in this type of soliton-mean flow problem. Ablowitz et al. \cite{Luo-stud-2023} studied the soliton-mean flow problem of the KdV equation with stepped initial values, and compared the results among soliton perturbation theory, the inverse scattering method and the Whitham modulation theory. Sprenger et al. \cite{EL-PRE-2018} developed the modulation theory of the soliton-mean flow problem of the defocusing NLS equation. In addition, the dynamical study of soliton-mean flow interaction is extended to nonconvex flow, in which a typical example is the mKdV equation \cite{EL-JFM-2021}.
\par
The paper is organized as follows. In Section \ref{sec:1}, we derive three Riemann invariants to describe the soliton-mean flow interaction, and then obtain the soliton modulation system. The overall framework of Whitham modulation theory is given in Section \ref{sec:2}, where some classical results are reviewed and the boundaries of the interaction region are characterized. Section \ref{sec:3} is the main content of this paper. It discusses in detail the interaction between solitons placed at different initial value positions and the mean flow, and provides the parameter conditions for soliton tunneling and soliton trapping. Some conclusions are summarized in Section \ref{sec:4}.

\section{Modulation dynamics of soliton-mean flow interaction}
\label{sec:1}

We now briefly describe the theory used to explain soliton-mean flow interactions, relying primarily on Whitham modulation theory, also known as solitonic dispersive hydrodynamics \cite{EL-PRL-2018}. Although solitons and mean flow vary on different scales, both of them are described by the same equations, which is a key feature of solitonic dispersive hydrodynamics. This section directly derives the modulation system that describes the soliton-mean flow interaction.

\subsection{Riemann invariants and solitonic modulation system}

Consider the general form of one-dimensional dispersive hydrodynamics, which is a hyperbolic conservation law model modified by the dispersion term, (see \cite{EL-PRL-2018},\cite{EL-JFM-2021})
$$u_t+F(u)_x=(D[u])_x,$$
taking the hydrodynamic flux function $F(u)=3u^2$, the term that differential operator acts on $D[u]=-u_{xx}$, then it corresponds to the KdV equation (\ref{KdV}).
\par
Substitute $u=\overline{u}+\varepsilon e^{i(kx-\omega t)}$ into KdV equation (\ref{KdV}) to obtain the dispersion relationship characterizing the KdV equation, i.e.,
$$
\omega(k,\overline{u})=6\overline{u}k-k^3, \quad k\in \mathbb{R},\quad k\neq 0.
$$
The dispersion relationship connects the local wave number $k$ with the local frequency $\omega$, which determines the kinematic properties of the wave train. In modulation theory, if the phase is labeled by $\theta=kx-\omega t$, then we have
$
k(x,t)=\frac{\partial \theta}{\partial x}, \quad \omega(x,t)=-\frac{\partial \theta}{\partial t}.
$
It is not difficult to find that there is a compatible equation connecting wave number and frequency, which is actually the conservation of waves, i.e.,
$$k_t+\omega_x=0.$$
In fact, solitonic dispersive hydrodynamics is closely related to the soliton limit of Whitham  modulation equation. The modulation system is expressed by physical wave parameters of the form
$$
\mathbf{u}_t+\mathbf{V}(\mathbf{u})\mathbf{u}_x=0,
$$
where for the KdV equation $\mathbf{u}\in \{(\overline{u},a,k)|\overline{u}\in \mathbb{R}, a>0, k \in \mathbb{R}\backslash \{0\} \}$ and the mean flow $\overline{u}$,
amplitude $a$, and wave number $k$ are all slowly varying functions of $x$ and $t$. The modulation period solution represented by them as parameters is the asymptotic solution of the system. In the soliton limit $k\rightarrow 0$, the modulation system admits the following exact reduction:
\begin{equation}
\begin{aligned}
\label{soliton-modulation-1}
&\overline{u}_t+6\overline{u}\overline{u}_x=0, \\
&a_t+g(a,\overline{u})\overline{u}_x+c(a,\overline{u})a_x=0.
\end{aligned}
\end{equation}
According to the conservation of waves, the equation below describes the soliton train
\begin{equation}
\label{soliton-modulation-2}
k_t+(c(a,\overline{u})k)_x=0,
\end{equation}
where $g(a,\overline{u})$ is a coupling function related to the background mean flow $\overline{u}$ and soliton amplitude $a$, and $c(a,\overline{u})$ represents the soliton amplitude-speed relation. Formula (\ref{soliton-modulation-1}) and formula (\ref{soliton-modulation-2}) constitute the solitonic modulation system.
\par
System (\ref{soliton-modulation-1}) can be rewritten into a diagonal form by defining $q=gd\overline{u}+(c-6\overline{u})da$
\begin{equation}
\label{modulation}
\left(
  \begin{array}{c}
    \overline{u} \\
    q \\
  \end{array}
\right)_t+\left(
            \begin{array}{cc}
              6\overline{u} & 0 \\
              0 & C(q,\overline{u}) \\
            \end{array}
          \right)\left(
  \begin{array}{c}
    \overline{u} \\
    q \\
  \end{array}
\right)_x=\left(
  \begin{array}{c}
    0 \\
    0 \\
  \end{array}
\right),
\end{equation}
where $C(q,\overline{u})\equiv c(q,\overline{u})$. The characteristic velocities of systems (\ref{soliton-modulation-1}) are $6\overline{u}$ and $C$, and the strict hyperbolicity requires $C\neq 6\overline{u}$.
It is pointed out here that $\overline{u}$ is ``hydrodynamic" Riemann invariant and satisfies the initial value condition, while $q$ is the ``solitonic" Riemann invariant, which exists as an adiabatic invariant \cite{EL-PRL-2018}.
\par
Introduce a conjugate (soliton) wave number $\widetilde{k}$, satisfying
$c(a,\overline{u})=\frac{\widetilde{w}(\widetilde{k},\overline{u})}{\widetilde{k}},$
where $\widetilde{w}(\widetilde{k},\overline{u})=-iw(i\widetilde{k},\overline{u})$. Then, we have
$\widetilde{k}^2=2a$, which implies that $\widetilde{k}\rightarrow 0$ is equivalent to $a\rightarrow 0$, i.e., $q=4\overline{u}$. This means that system (\ref{modulation}) degenerates into a single hyperbolic equation $\overline{u}_t+6\overline{u}\overline{u}_x=0$. It happens to be the KdV equation after removing the dispersion term, also known as the Hopf equation.
\par
Following the work of \cite{EL-Chaos-2005}, the adiabatic invariants can be obtained directly from the following equation
$$
\frac{d\widetilde{k}}{d\overline{u}}=
\frac{\widetilde{w}_{\overline{u}}}{6\overline{u}-\widetilde{w}_{\widetilde{k}}}.
$$
After integral operation, it follows
\begin{equation}
\label{q}
q(a,\overline{u})=4\overline{u}+2a,
\end{equation}
and
\begin{equation}
\label{C}
C(q,\overline{u})=2\overline{u}+q.
\end{equation}
It is desirable to write equation (\ref{soliton-modulation-2}) in diagonal form, so define
\begin{equation*}
p(q,\overline{u})=\mathrm{exp}(-\int_{\overline{u}_0}^{\overline{u}}\frac{C_u(q,u)}{F^{'}(u)-C(q,u)}du).
\end{equation*}
Reminding $C(q,u)=2u+q$ and $F^{'}(u)=6u$, then $p$ is determined by
\begin{equation}
p(q,\overline{u})=\mathrm{exp}\left\{-\int_{\overline{u}_0}^{\overline{u}}\frac{2}{6u-(2u+q)}du\right\}
=\frac{1}{\sqrt{(4\overline{u}-q)}},
\end{equation}
where it is convenient to take $\overline{u}_0=\frac{1-q}{4}$. In fact, if $q$ is a constant, this is also consistent with the soliton-mean flow interaction we consider. The degraded system (\ref{modulation}) combined with equation (\ref{soliton-modulation-2}) still results in a hyperbolic system. In other words, the following equation can be used to replace equation (\ref{soliton-modulation-2}) with
$$
(kp)_t+C(q,\overline{u})(kp)_x=0.
$$
In fact, $r=kp(q,\overline{u})$ is regarded as an adiabatic invariant, which is the third Riemann invariant \cite{EL-PRL-2018}. Notice that the determination of Riemann invariants is very important. The constants of $q$ and $r$ directly lead to the transmission condition. The trajectory of the soliton after interacting with the mean flow also depends on this.

\subsection{Transmission condition and phase relation}
\par
The KdV equation (\ref{KdV}) allows the existence of soliton solutions in the following form
\begin{equation}
\label{exact-soliton}
u=\overline{u}+a\mathrm{sech}^2\left(\sqrt{\frac{a}{2}}(x-(2a+6\overline{u})t-x_0)\right),
\end{equation}
in which the soliton amplitude is $a$, and the velocity of the soliton propagating to the right on background $\overline{u}$ is related to its amplitude and background and can be expressed as $2a+6\overline{u}$.
Scale separation requires that rarefaction wave (RW) or dispersive shock wave (DSW) show an extended fan region or oscillation region that increases with time and the width is much larger than the soliton width. Inspired by exact soliton solution (\ref{exact-soliton}), the initial amplitude of the trial soliton is $a_0=a(x,0)$ located at $x=x_0$.
\par
The soliton modulation system (\ref{soliton-modulation-1}) and (\ref{soliton-modulation-2}) can be described by a simple wave (to be justified), which means that there is only one Riemann invariant that is not a constant, which is $u$ that satisfies the initial value. That is to say for the three Riemann invariants introduced in the previous section, $\overline{u}$ satisfies the initial condition, and $q(a,\overline{u})$ and $r$ serve as global adiabatic invariants. According to this assertion, for all $x$, the initial amplitude $a(x,0)$ and wave field $k(x,0)$ of the soliton can be defined as
\begin{equation*}
\begin{aligned}
&q(a(x,0),\overline{u}(x,0))=q_0 \equiv q(a_0,\overline{u}(x_0,0)),\\
&k(x,0)p(q_0,\overline{u}_0(x))=k(x_0,0)p(q_0,\overline{u}_0(x_0)), \quad i.e. \quad k(x,0)=\frac{k_0p_0}{p(q_0,\overline{u}_0(x))},
\end{aligned}
\end{equation*}
where $0<k_0=k(x_0,0), p_0=p(q_0,\overline{u}_0(x_0))\ll 1.$
\par
Since the region evolved from discontinuity is very complicated, the interaction between the soliton and this region can be expressed through the changes before and after. Corresponding to the initial value problem (\ref{Initial condition}), the initial wave field is described by
\begin{equation}
a(x,0)=\left\{
\begin{aligned}
&a_L,  &x\leq 0,\\
&a_M,  &0<x<l,\\
&a_R,  &x\geq l,
\end{aligned}
\right.
\quad
k(x,0)=\left\{
\begin{aligned}
&k_L,  &x\leq 0,\\
&k_M,  &0<x<l,\\
&k_R,  &x\geq l.\\
\end{aligned}
\right.
\end{equation}
\par
The invariance of the Riemann invariants $q,r$ directly derives the transmission condition and phase relation of the soliton mean flow problem below
\begin{equation}
q(a_L,u_L)=q(a_R,u_R),
\label{transmission}
\end{equation}
\begin{equation}
\label{invariance}
k_Lp(q_0,u_L)=k_Rp(q_0,u_R).
\end{equation}
\begin{figure}
\centering
\includegraphics[width=11.0cm]{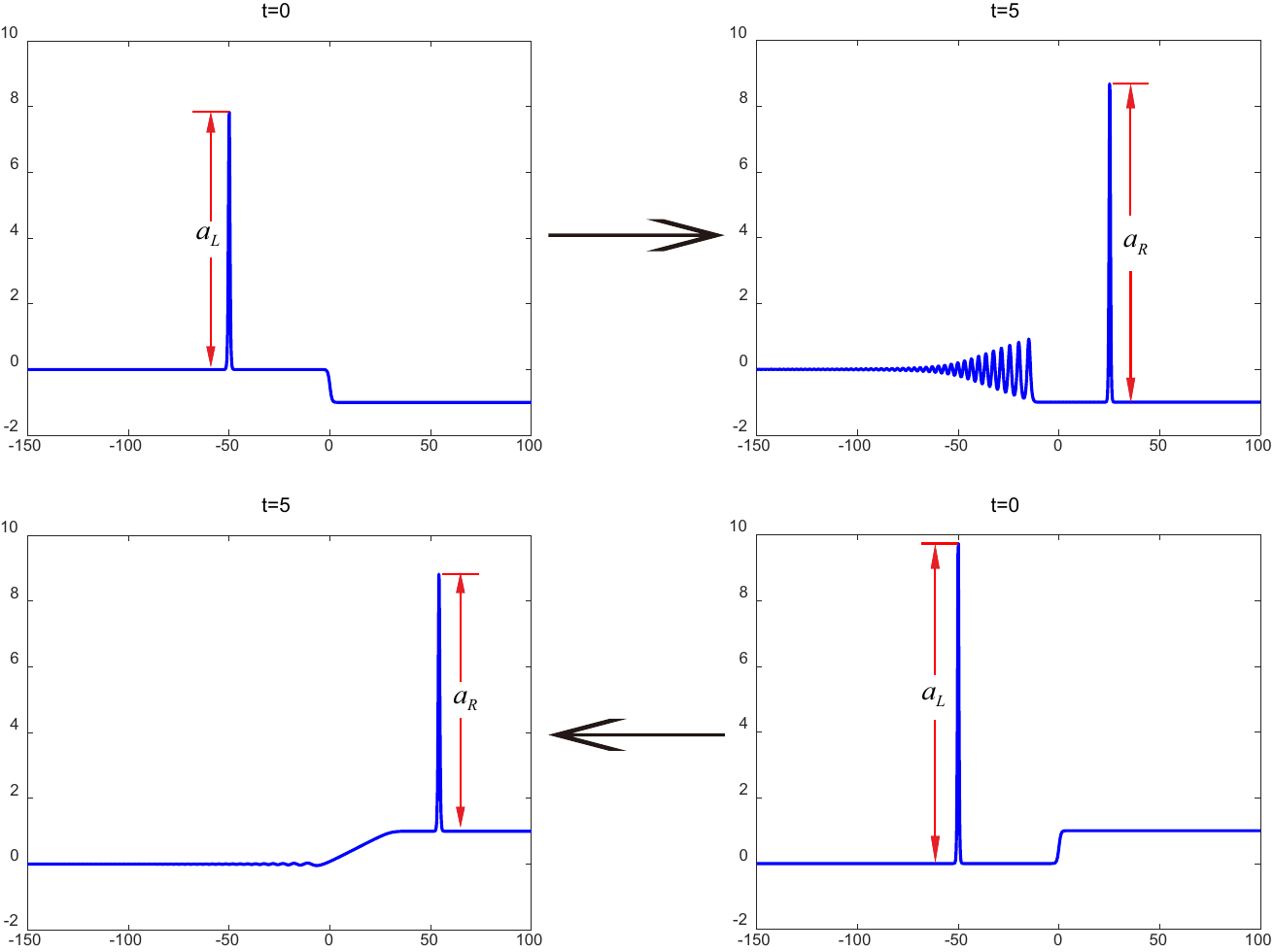}
\caption{{\protect\small Graphical depictions of time reciprocity.}}
\label{timereverbility-fig}
\end{figure}
\par
Introduce $\delta$ to characterize the phase shift of soliton
$$
\delta=x_+-x_-,
$$
that is, the phase difference between the phase position succeeding $(x_{+})$ and preceding $(x_{-})$ the tunneling region. Morover, let $x=x_{\mp}+c(a,\overline{u})t$ represent the soliton trajectories before and after soliton tunneling, respectively. There is a fact that needs to be explained. The step initial data will generate RW and DSW, or the mean flow depends on the order of $u_L$ and $u_R$. Hydrodynamic reciprocity guarantees that we can still find a simple wave solution for soliton-DSW modulation, ensuring that the soliton modulation system (\ref{soliton-modulation-1}) and (\ref{soliton-modulation-2}) are still valid. In fact, the time reversibility of the KdV equation guarantees this hydrodynamic reciprocity, as shown graphically in Fig. \ref{timereverbility-fig}. If the soliton that passes through the DSW is used as the initial state that interacts with the RW, the soliton that passes through the RW maintains the same properties as the soliton that does not pass through the DSW.
\par

%

\section{Mean flow: Whitham modulation theory}
\label{sec:2}

The Whitham modulation theory of KdV equation has been relatively completely explained in the classical paper published in 1980 \cite{FFM-CPAM-1980}. RW and DSW can be described by self-similar solutions in Gurevich-Pitaevskii (GP) problem \cite{Gurevich-1974} with a single discontinuous piecewise constant initial data at the origin. In particular, El and Grimshaw \cite{EL-Chaos-2002} studied the Whitham modulation theory of the small dispersion KdV equation with initial data in the form of rectangular pulses. We will briefly review some important results in this work, which are an indispensable part of the theory of soliton-mean flow modulation.

\subsection{Periodic solution}

Conventionally, take $\xi=x-Vt$ to obtain the traveling wave solution of the KdV equation, which is described by a third-order ODE
\begin{equation*}
(u_\xi)^2=G(u)=-2(u-u_1)(u-u_2)(u-u_3),
\end{equation*}
where $G(u)=-(2u^3-Vu^2-2Au-2B)$ with constants
\begin{equation*}
V=2(u_1+u_2+u_3),\quad
A=-(u_1u_2+u_1u_3+u_2u_3),\quad
B=u_1u_2u_3.
\end{equation*}
Assume that the order relationship of the three roots is $u_1\leq u_2\leq u_3$, then the periodic solution is represented by the Jacobian elliptic function
\begin{equation}
\label{periodic-solution}
u(x,t)=u_2+(u_3-u_2)\mathrm{cn}^2(\sqrt{2(u_3-u_1)}(\xi-\xi_0);m),
\end{equation}
where $m=\frac{u_3-u_2}{u_3-u_1}$ is modulus of Jacobian elliptic function. The wavelength is defined as
\begin{equation}
L=2\int_{u_2}^{u_3}\frac{d u}{\sqrt{-G(u)}}=\frac{2\sqrt{2}K(m)}{\sqrt{u_3-u_1}}.
\end{equation}
Here the wave number is $k=\frac{2\pi}{L}$ and the frequency is $w=kV$.
When $m\rightarrow0$, it is harmonic wave with vanishing amplitude, and the solution degenerates into
$$
u_h(x,t)\approx u_3-(u_3-u_2)\mathrm{sin}^2(k_h(x-V_ht)),
$$
where $V_h=6u_3-k_0^2$ conforms to the dispersion relationship of the KdV equation, when $\overline{u}=u_3$.
We pay more attention to the soliton limit, i.e., when $m\rightarrow1$, the periodic solution degenerates into a solitary wave
$$
u_s(x,t)=u_1+(u_3-u_1)\mathrm{sech}^2(\sqrt{2(u_3-u_2)}(x-V_st)),
$$
where $V_s=6u_1+2(u_3-u_1)=6\overline{u}+2a$. Since when $|x|\rightarrow \infty$, $u_s(x,t)\rightarrow u_1$, it implies that the background wave is $u_1$. According to the restrictions on the background wave by the initial value condition (\ref{Initial condition}) and the order relationship of $u_i$ we previously assumed, it can be seen that this is a soliton traveling wave solution moving to the right.
\par
Introducing Riemann invariants
$$
\lambda_1=\frac{u_1+u_2}{2},\quad \lambda_2=\frac{u_1+u_3}{2},\quad \lambda_3=\frac{u_2+u_3}{2},
$$
then according to the order relationship of $u_i~(i=1,2,3)$, the order $\lambda_3\geq \lambda_2 \geq \lambda_1$ is established. Then the periodic solution (\ref{periodic-solution}) can be expressed as Riemann invariant and rewritten as
\begin{equation*}
u=\lambda_1-\lambda_2+\lambda_3+2(\lambda_2-\lambda_1)\mathrm{cn}^2(2\sqrt{(\lambda_3-\lambda_1)}(x-Vt)+\xi_0,m),
\end{equation*}
where $m=\frac{\lambda_2-\lambda_1}{\lambda_3-\lambda_1}$. At this time, the harmonic limit $\lambda_2\rightarrow\lambda_1$ corresponds to small amplitude state
\begin{equation*}
u\sim\lambda_3+2(\lambda_2-\lambda_1)\mathrm{cos}^2(2\sqrt{(\lambda_3-\lambda_1)}\xi+\xi_0),
\end{equation*}
and the soliton limit $\lambda_2\rightarrow\lambda_3$ corresponds to soliton state
\begin{equation*}
u=\lambda_1+2(\lambda_{23}-\lambda_1)\mathrm{sech}^2(2\sqrt{(\lambda_{23}-\lambda_1)}(x-(2\lambda_1+4\lambda_{23})t)).
\end{equation*}

\subsection{Modulation equations}

\begin{figure}
\centering
\includegraphics[width=12.0cm]{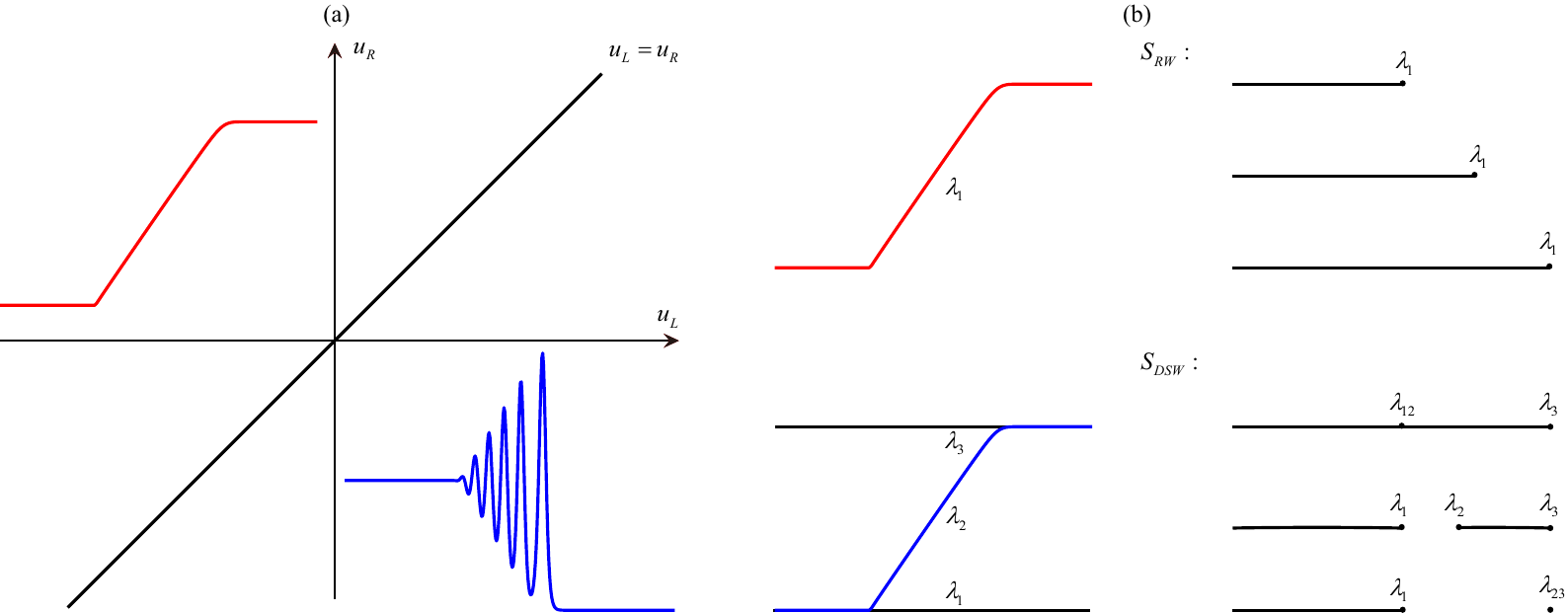}
\caption{{\protect\small Classification of solutions to the dispersion Riemann problem of KdV equation.}}
\label{basic-class-fig}
\end{figure}

The Riemann invariant $\lambda_i~(i=1,2,3)$ can approximately construct the solution to the initial value problem by satisfying the modulation equation, which is a key point in the modulation theory. In order to facilitate the subsequent description, the wavelength $L$ is rewritten here using Riemann invariants,
\begin{equation}
L=\frac{2K(m)}{\sqrt{\lambda_3-\lambda_1}}.
\end{equation}
The one-phase KdV-Whitham modulation system is allowed to be shown in a diagonal form,
\begin{equation}\label{KdV-Whitham-E}
\frac{\partial \lambda_i}{\partial t}+v_i(\lambda)\frac{\partial \lambda_i}{\partial x}=0,\quad ~i=1,2,3,
\end{equation}
where the Whitham velocity is given in terms of group velocity
\begin{equation}
\label{vj}
v_j=\frac{\partial_j w}{\partial_j k}=V-2\frac{L}{\partial_j L},
\end{equation}
and specifically
\begin{equation}
\begin{aligned}
&v_1=2(\lambda_1+\lambda_2+\lambda_3)-4(\lambda_2-\lambda_1)\frac{K(m)}{K(m)-E(m)},\\
&v_2=2(\lambda_1+\lambda_2+\lambda_3)-4(\lambda_2-\lambda_1)\frac{(1-m)K(m)}{E(m)-(1-m)K(m)},\\
&v_3=2(\lambda_1+\lambda_2+\lambda_3)+4(\lambda_3-\lambda_2)\frac{K(m)}{E(m)},
\end{aligned}
\end{equation}
where the modulus $m=\frac{\lambda_2-\lambda_1}{\lambda_3-\lambda_1}$ $(0\leq m\leq 1)$, $K(m)$ and $E(m)$ are complete elliptic integrals of the first and second kind, respectively.
\begin{figure}
\centering
\includegraphics[width=11.0cm]{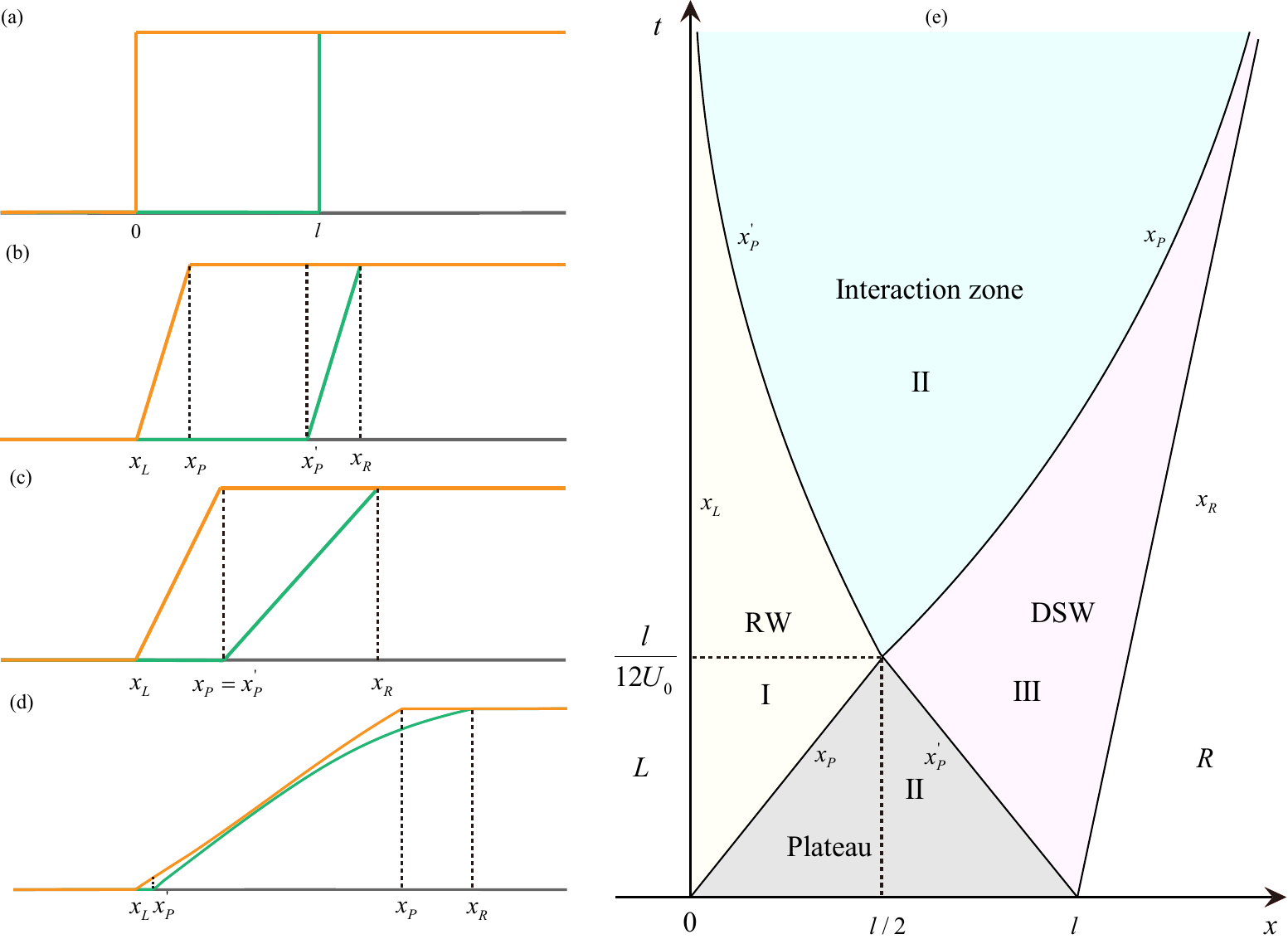}
\caption{{\protect\small (a)-(d) Evolution process of Riemann invariants; (e) Evolution of the initial value of box type in the $(x,t)$ plane.}}
\label{lambda-whitham-fig}
\end{figure}
\par
For a single discontinuous step-like initial value, the solution to the GP problem \cite{AMK-1997,Gurevich-1974} of the KdV equation has been classified into RW and DSW, which is an undoubted result (see Fig. \ref{basic-class-fig}). The step-up initial value corresponds to RW, and conversely, the step-down initial value evolves to DSW. The changes in spectrum corresponding to RW and DSW are also shown in Fig. \ref{basic-class-fig}.  For one-phase spectrum $S_{DSW}$, a band shrinks to a point at the limit $\lambda_2\rightarrow \lambda_3$, which corresponds to the soliton limit. This is the degeneration of the one-phase solution and can also be regarded as the a semi-infinite band of the zero-phase solution and a characteristic spectrum point. This is instructive for the study of soliton mean flow interactions.
\begin{figure}
\centering
\includegraphics[width=7.0cm]{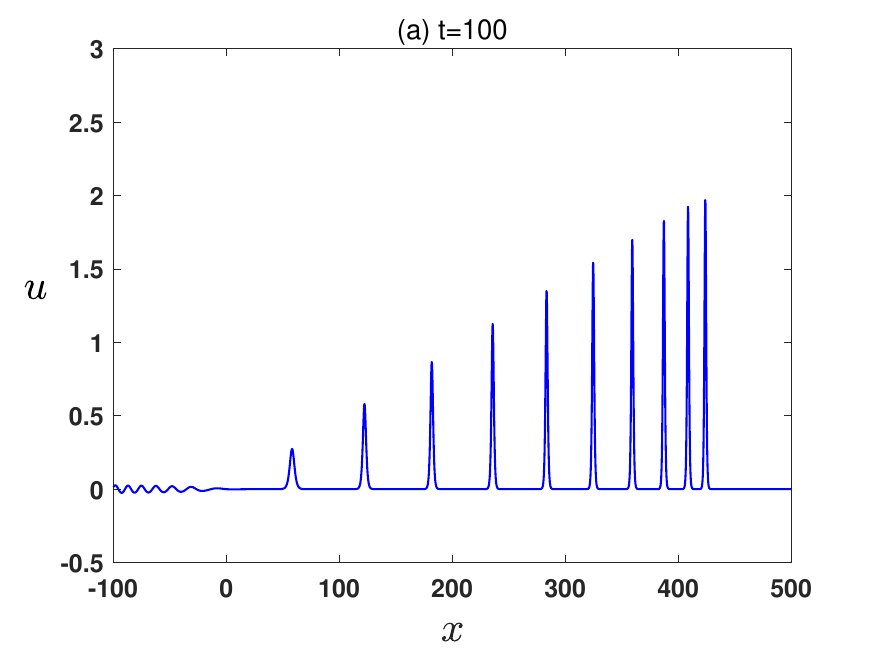}
\includegraphics[width=7.0cm]{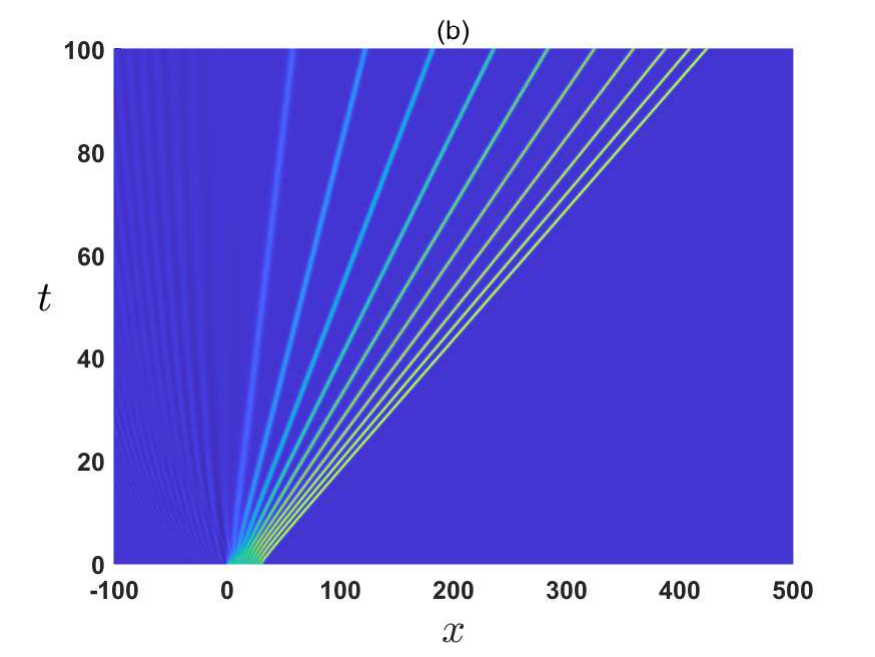}
\caption{{\protect\small (a) Evolution results of solutions to the KdV equation with box initial value over a long period of time; (b) Evolution result of the initial value of box type in the $(x,t)$ plane. The parameter selection in (\ref{Initial condition}) is $l=10\pi, U_0=1$.}}
\label{box-whitham-fig}
\end{figure}
\par
Focus on the initial value (\ref{Initial condition}), which is different from the single discontinuous initial value type in \cite{Luo-stud-2023}. The initial value condition in the form of an obstacle will cause the RW region and the DSW region to interact to form a region in which a limited number of soliton train are arranged, called the interaction zone. The varying of Riemann invariants are shown in Fig. \ref{lambda-whitham-fig}(a)-(d). The emergence of soliton train can also be explained from the perspective of Riemann invariants. The limit state $\lambda_2\rightarrow\lambda_3$ corresponds to the soliton limit $m\rightarrow1$. As time develops, the $(x,t)$ plane is divided into six regions during the long-term evolution of the initial data (\ref{Initial condition}). Fig. \ref{box-whitham-fig} displays the direct numerical simulation of the KdV equation with the box initial value. After the time passes, the soliton region is very obvious and almost occupies the entire wave region.
In fact, the number of soliton $N$ is equal to the number of eigenvalues of the corresponding linear spectrum problem, and can be calculated, which was first obtained by Karpman \cite{Karpman-1967}
\begin{equation}
\label{soliton-number}
N\approx \frac{1}{\pi}\int_{-\infty}^{\infty}\sqrt{u_0(x)}dx.
\end{equation}
Observe the equation (\ref{soliton-number}) and notice that the number of solitons is independent of the evolution of time. However, as the region lengthens, the width of each soliton in the soliton train also becomes larger. For simplicity of calculation, take $l=10\pi, U_0=1$. According to the equation (\ref{soliton-number}), the number of solitons is ten, which is consistent with the numerical simulation results in Fig. \ref{box-whitham-fig}.
\par
Frankly speaking, the calculation of the boundary is of great significance for determining the soliton trajectory, and the boundary of the region appearing here can indeed be obtained. When there is no interaction zone, that is, within a limited time, the plane is divided into five regions, and the three plane wave regions sandwich the RW and DSW regions. The boundaries are as follows:
\begin{equation}
\begin{aligned}
\label{before-bound}
&x_L=v_3(0,0,\lambda_3)t=0,\quad
x_P=v_3(0,0,\lambda_3)t=6U_0t,\\
&x_P^{'}=l+v_2(0,0,\lambda_3)t=l-6U_0t,\quad
x_R=l+v_2(0,\lambda_3,\lambda_3)t=l+4U_0t.
\end{aligned}
\end{equation}
\par
When $x_P=x_P^{'}$, the plateau disappears exactly, and next the interaction zone will appear. Observing the equation (\ref{before-bound}), it is not difficult to find that there is always such a moment when the plateau region begins to disappear. In fact, this moment depends on the setting of the initial value, $T=\frac{l}{12U_0}$. We focus on the boundaries of the interaction region. As for $x_P^{'}$, it is clear that $\lambda_3(x_P^{'})$ satisfies the Hopf equation, and
$$
\lambda_2(x_P^{'})=\lambda_1(x_P^{'})=0.
$$
As for $x_P$, continuity guarantees
$$
\lambda_2(x_P-0)=\lambda_2(x_P+0),\quad \lambda_3(x_P)=1,\quad \lambda_1(x_P)=0.
$$
Now $\lambda_1$ is always a constant, and only the Whitham equations corresponding to $\lambda_2$ and $\lambda_3$ are considered, i.e.,
$$
\begin{aligned}
&\frac{\partial \lambda_2}{\partial t}+v_2(0, \lambda_2, \lambda_3)\frac{\partial \lambda_2}{\partial x}=0, \\
&\frac{\partial \lambda_3}{\partial t}+v_3(0, \lambda_2, \lambda_3)\frac{\partial \lambda_3}{\partial x}=0.
\end{aligned}
$$
By hodograph transform $x-v_j t=w_j~ (j=2,3)$, it can be rewritten as
\begin{equation}
\label{hodograph-system}
\frac{\partial_2 w_3}{w_2-w_3}=\frac{\partial_2 v_3}{v_2-v_3},\quad
\frac{\partial_3 w_2}{w_3-w_2}=\frac{\partial_3 v_2}{v_3-v_2}.
\end{equation}
After hodograph transform, the purpose is transformed from determining the boundary conditions of the Riemann invariant to determining the boundary conditions of the function $w(\lambda_2,\lambda_3)$ under the new plane
\begin{equation}
\label{hodograph-BC}
w_3(0,\lambda_3)=0,\quad w_2(\lambda_2,1)=l.
\end{equation}
Note that $w_j$ and $v_j$ are symmetrical positions in the system (\ref{hodograph-system}) after hodograph transform. Here $w_j$ can be expressed as
\begin{equation}
\label{wj}
w_j=\frac{\partial_j(kf)}{\partial_j k}=f-\frac{L}{\partial_j L}\partial_j f,
\end{equation}
where $\partial_j\equiv \frac{\partial}{\partial \lambda_j}$, and $f=f(\lambda_2, \lambda_3)$ is an unknown function. The motivation is to express $w_j$ as the form of $v_j$ in (\ref{vj}).
Then the Euler-Darboux-Poisson equation for $f$ is derived
\begin{equation}
2(\lambda_2-\lambda_3)\partial_{23}^2f=\partial_2 f-\partial_3 f.
\end{equation}
Rewrite boundary condition (\ref{hodograph-BC}) as
\begin{equation}
\label{f-BC}
f(0,\lambda_3)+2\lambda_3\partial_3f(0,\lambda_3)=0,\quad f(\lambda_2,1)-\frac{K(\lambda_2)}{K^{'}(\lambda_2)}\partial_2f(\lambda_2,1)=l.
\end{equation}
Integrate the equation (\ref{f-BC}), then we have
\begin{equation}
f(\lambda_2,1)=C_1K(\lambda_2)+l,\quad f(0,\lambda_3)=\frac{C_2}{\sqrt{\lambda_3}}.
\end{equation}
Here $C_1$ and $C_2$ are arbitrary integration constants. For simplicity, take $C_1=0$. Next, rewrite the Euler-Darboux-Poisson equation into the form Euler-Poission operator
$$
E_{ij}f=0,
$$
where $E_{ij}=\partial_{ij}^2-\frac{\partial_{i}-\partial_{j}}{2(\lambda_i-\lambda_j)}$.
This equation has a solution of the following form \cite{Tricomi-1961}
\begin{equation}
f=\int_0^{\lambda_2}\frac{\phi_1(\beta)d \beta}{\sqrt{\beta(\lambda_3-\beta)(\beta-\lambda_2)}}+
\int^1_{\lambda_3}\frac{\phi_2(\beta)d \beta}{\sqrt{\beta(\beta-\lambda_3)(\beta-\lambda_2)}},
\end{equation}
where $\phi_1(\beta)$ and $\phi_2(\beta)$ are arbitrary functions.
\par
Combining boundary conditions and inverse Abel transform, $\phi_1$ and $\phi_2$ can be determined by
\begin{equation}
\phi_1(\beta)=\frac{l\sqrt{\beta-1}}{\pi},\quad \phi_2\equiv0.
\end{equation}
Then it follows
\begin{equation}
f(\lambda_2,\lambda_3)=\frac{2l}{\pi\sqrt{(1-\lambda_2)\lambda_3}}{\prod}_{1}
\left(\frac{\lambda_2}{\lambda_2-1},
\frac{(1-\lambda_3)\lambda_2}{(1-\lambda_2)\lambda_3}\right),
\end{equation}
where $\prod_1$ is the complete elliptic integral of the third kind.
\par
The dependence of the Riemann invariant $\lambda$ on $x, t$ is reconstructed through equation (\ref{wj}) and hodograph transform. With the above preparation, it is ready to return to the boundary of the interaction region that we are concerned about. The boundary $x_P^{'}$ connected to the rarefaction wave satisfies
$$
\frac{d x_P^{'}}{d t}=v_1(0,0,\lambda_3)=v_2(0,0,\lambda_3)=-6\lambda_3.
$$
where $\lambda_3$ must satisfies the Hopf equation. Combined with the information at the initial moment $T$ when the interaction occurs, one can get
\begin{equation}
\label{BC-IZ-left}
x_P^{'}=\frac{l^2}{24U_0 t}.
\end{equation}
The boundary $x_P$ connecting the interaction region and the dispersive shock wave region is
$$
x_P
=(2U_0(1+m)-\frac{4mU_0(1-m)}{\mu(m)-(1-m)})t+l,
$$
where $\mu(m)=\frac{E(m)}{K(m)}$ and $m=\frac{\lambda_2}{U_0}$. Note that when $m\rightarrow 1$, one has $x_P\rightarrow 4U_0t+l$, which means that $x_P$ gradually approaches $x_R$.
\par
In fact, when $m\rightarrow1$ (i.e. $\lambda_2\rightarrow\lambda_3$), the Whitham modulation system degenerates to
\begin{equation}
\begin{aligned}
&\frac{\partial \lambda_1}{\partial t}+6\lambda_1\frac{\partial \lambda_1}{\partial x}=0, \\
&\frac{\partial \lambda_3}{\partial t}+(2\lambda_1+4\lambda_3)\frac{\partial \lambda_3}{\partial x}=0.
\end{aligned}
\end{equation}
Denote the background wave as $\overline{u}=u_1=\lambda_1$ and the soliton amplitude as $a=u_3-u_1=2(\lambda_3-\lambda_1)$. Then the correspondence with the soliton modulation system can be achieved at
\begin{equation}
\label{re-whitham}
\begin{aligned}
&\frac{\partial \overline{u}}{\partial t}+6\overline{u}\frac{\partial \overline{u}}{\partial x}=0, \\
&\frac{\partial a}{\partial t}+(6\overline{u}+2a)\frac{\partial a}{\partial x}+2a\frac{\partial \overline{u}}{\partial x}=0,
\end{aligned}
\end{equation}
where the coupling function $g(a,\overline{u})=2a$, and $C(\overline{u},a)=6\overline{u}+2a$.
Based on $q=4\overline{u}+2a$, we can get
\begin{equation}
\label{q-soliton}
\frac{\partial q}{\partial t}+(2\overline{u}+q)\frac{\partial q}{\partial x}=0.
\end{equation}
This equations (\ref{q-soliton}) and (\ref{re-whitham}) are actually the modulation system (\ref{modulation}).
\par
The interaction between soliton and mean flow will be described by Whitham modulation theory. Theoretically, the degenerate one-phase Whitham modulation equation is very effective in describing the soliton-RW interaction, and the soliton-DSW interaction can use the method of the degenerate two-phase Whitham modulation equation.
Based on the pioneering work of Flaschka, Forest, and McLaughlin \cite{FFM-CPAM-1980}, the Whitham modulation system describing the N-phase solution of the KdV equation can be obtained. The results of the one-phase have been given before. As preliminary work, the $N=2$ genus Whitham modulation equation is repeated below
\begin{equation}\label{N2-Whitham}
\frac{\partial \lambda_k}{\partial t}+v_k^{(2)}\frac{\partial \lambda_k}{\partial x}=0,\quad k=1\ldots 5.
\end{equation}
Introduce the following integral representation
\begin{equation}
I_i^j(\lambda_k)=\int_{\lambda_{2i-1}}^{\lambda_{2i}}\frac{(\lambda_k-\mu)\mu^j d\mu}{P(\mu)},
\end{equation}
where $P(\mu)=\sqrt{(\mu-\lambda_1)(\mu-\lambda_2)(\mu-\lambda_3)(\mu-\lambda_4)(\mu-\lambda_5)}$, then the Whitham velocities in (\ref{N2-Whitham}) can be expressed as
\begin{equation}
v_k^{(2)}=-6\Sigma_{k=1}^5 \lambda_k+12\lambda_k+12\frac{I_2^2(\lambda_k)I_1^0(\lambda_k)-I_2^0(\lambda_k)I_1^2(\lambda_k)}
{I_2^1(\lambda_k)I_1^0(\lambda_k)-I_2^0(\lambda_k)I_1^1(\lambda_k)}.
\end{equation}


\section{Modulation theory for soliton$-$mean flow interaction}
\label{sec:3} In fact, there are only two results of the long-term evolution of soliton, trapping or tunneling. Placing the initial soliton at different positions may lead to different results during the dynamic evolution process, which is directly related to Whitham velocities. Below we discuss three types of initial soliton positions, which are $x_0<0$, $0<x_0<l$, and $l<x_0$. Before the formal discussion, let us do some preparatory work and focus on the interaction of this trial soliton with RW and DSW, respectively.
\par
\begin{figure}
\centering
\includegraphics[width=6cm]{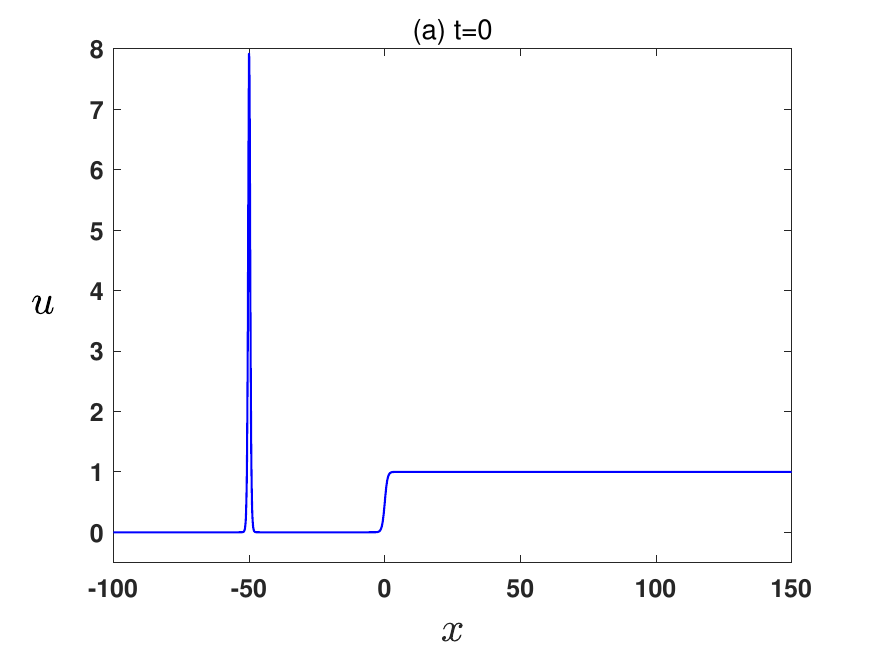}
\includegraphics[width=6cm]{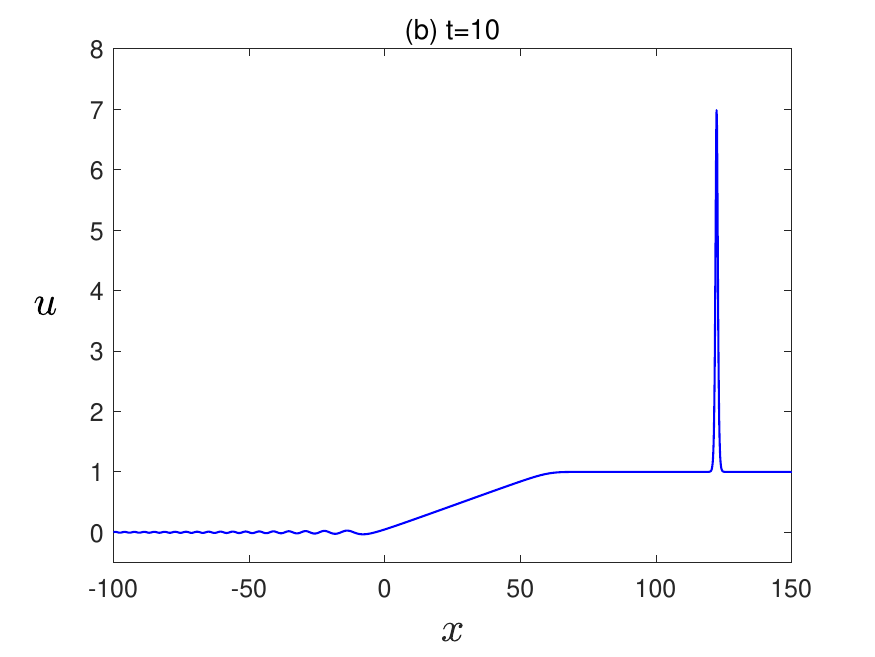}\\
\includegraphics[width=6cm]{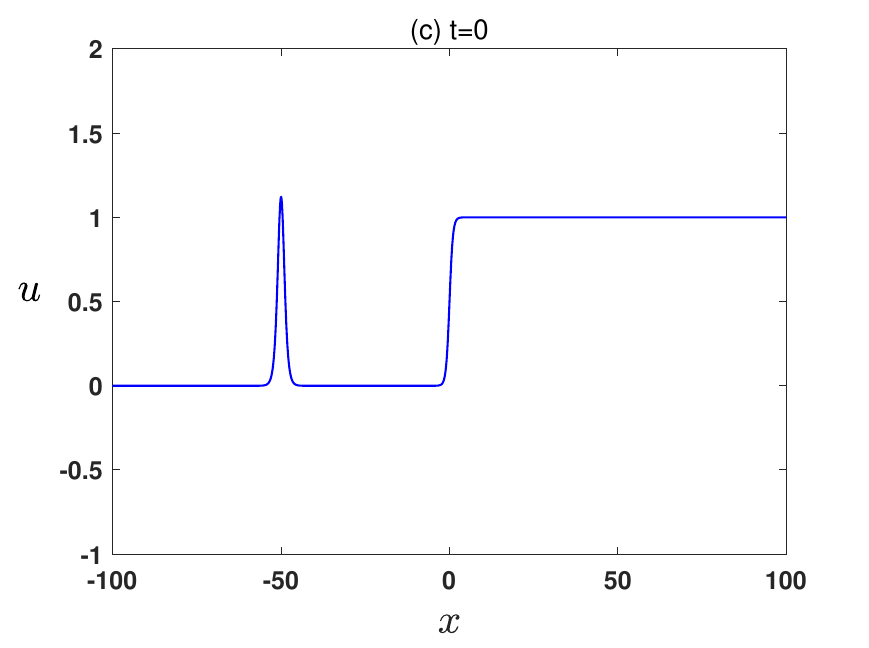}
\includegraphics[width=6cm]{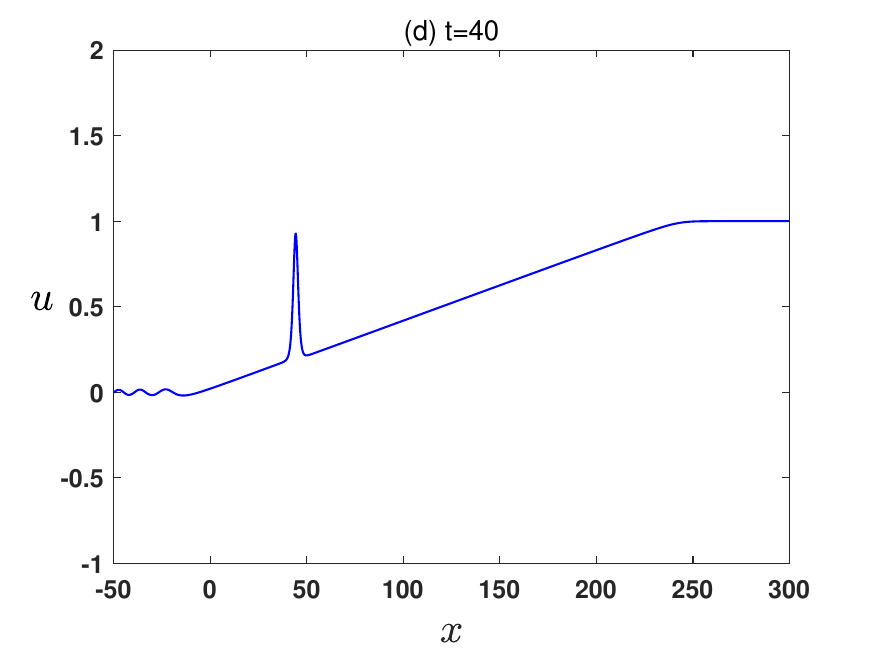}
\caption{{\protect\small Dynamical evolutions of soliton-RW interactions; (a)-(b) soliton tunneling; (c)-(d) soliton trapping.}}
\label{s_RW_tunnel_trapping-fig}
\end{figure}

{\bf Soliton-RW:} For the basic rarefaction wave structure, the left boundary is fixed at $0$. The right boundary can be calculated from the Whitham velocity. It is not difficult to find that if the initial soliton is located on the right side of the discontinuity, that is, $x_0>0$, its speed is greater than the right boundary speed of the rarefaction wave, and there will be no interaction, which is not what we are concerned about. If the soliton is placed on the left side $(x_0<0)$, the transmission condition can be obtained by analyzing the global invariant $q$, that is
\begin{equation}
a_{L}>a_{cr}=2(\overline{u}_R-\overline{u}_L),
\end{equation}
where $\overline{u}_R$ and $\overline{u}_L$ are the left and right backgrounds of the step-like initial value. Fig.\ref{s_RW_tunnel_trapping-fig} (a)-(b) shows the process of soliton crossing the RW. The process of interaction between soliton and RW can be regarded as an elongated process, and the amplitude becomes smaller. The position of the soliton after crossing can be approximately estimated according to the transmission condition (\ref{transmission}) and phase condition (\ref{invariance}). Fig.\ref{s_RW_tunnel_trapping-fig} (c)-(d) is the process of soliton trapping in RW.
\par
\par
\begin{figure}
\centering
\includegraphics[width=6cm]{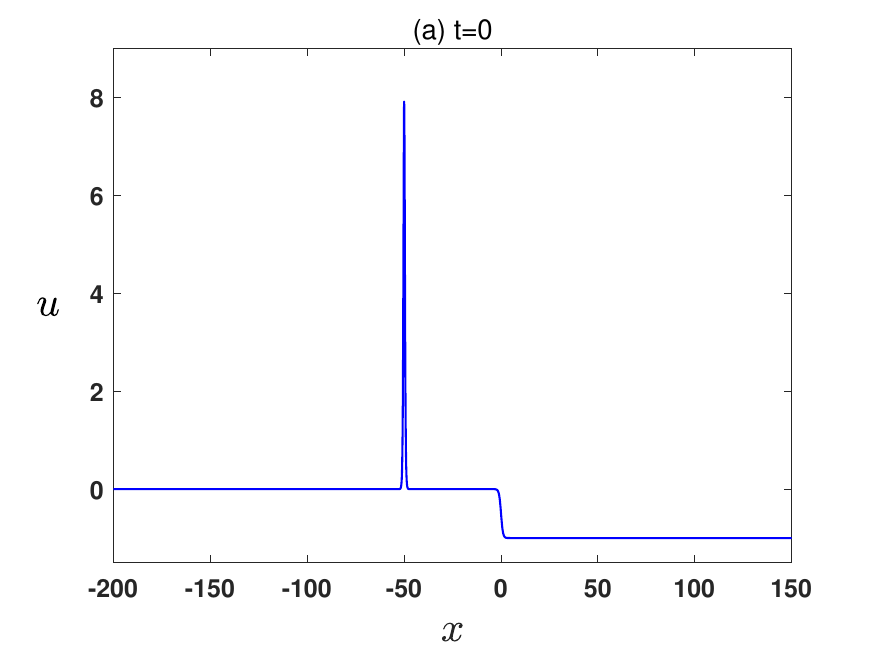}
\includegraphics[width=6cm]{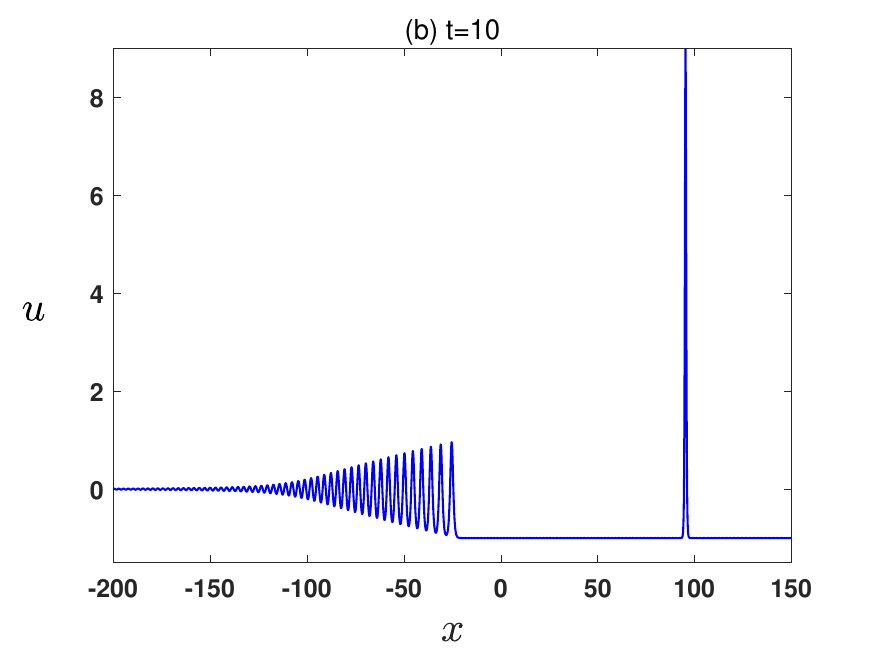}\\
\includegraphics[width=6cm]{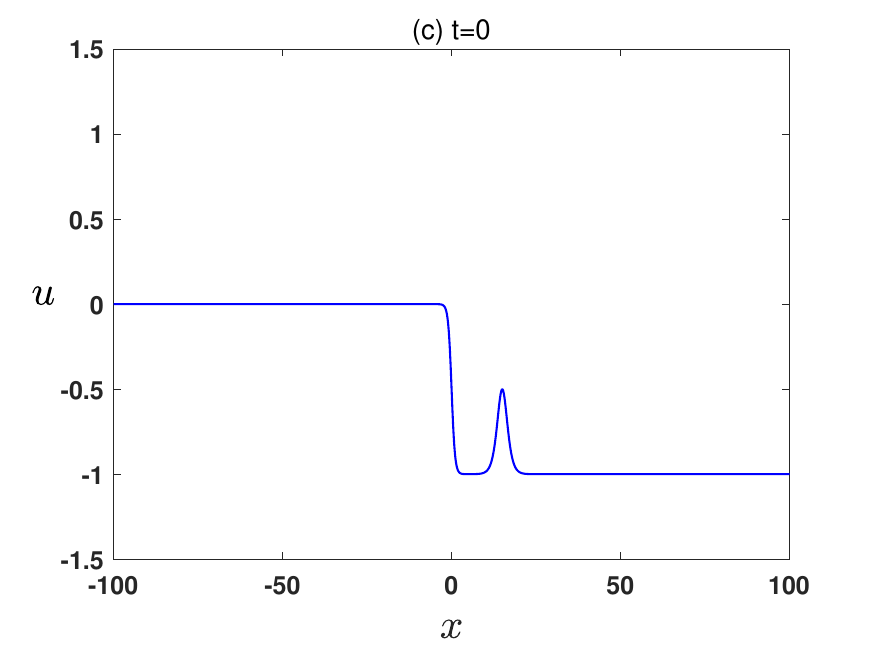}
\includegraphics[width=6cm]{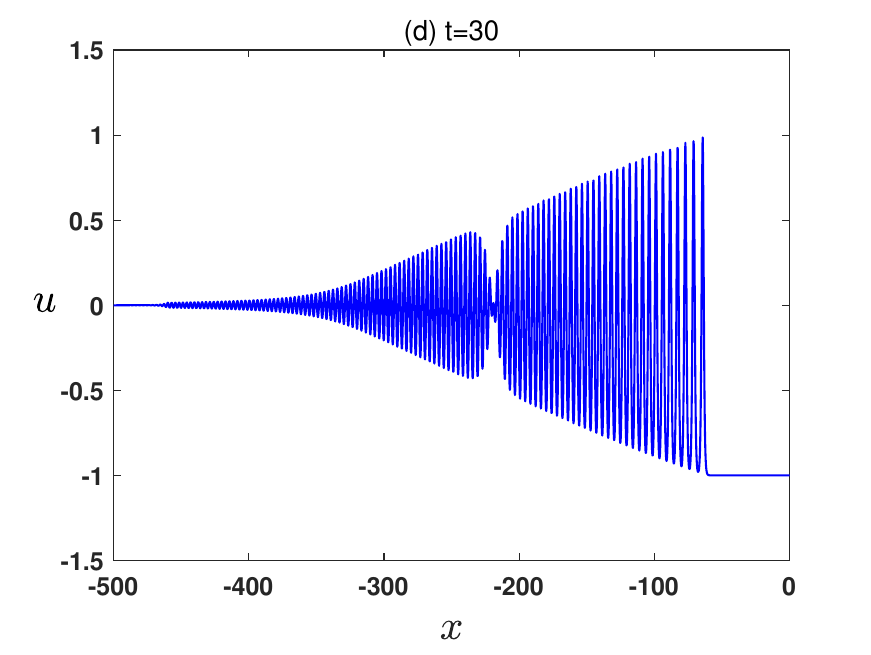}
\caption{{\protect\small Dynamical evolutions of soliton-DSW interactions; (a)-(b) soliton tunneling; (c)-(d) soliton trapping.}}
\label{s_DSW_tunnel_trapping-fig}
\end{figure}

{\bf Soliton-DSW:} Hydrodynamic reciprocity guarantees that the soliton-DSW interaction satisfies the same transport and phase conditions as the soliton-RW interaction, which relies on the time reversibility of the KdV equation (\ref{KdV}) \cite{EL-PRL-2018}. In fact, if the soliton is placed on the left side, the soliton will always interact non-trivially with the DSW and propagate, that is, soliton tunneling \cite{Luo-stud-2023}. The soliton placed on the right side will be trapped inside the DSW region if and only if $0<a_-<2U_0$, because the speed of the soliton propagating to the right is smaller than the soliton boundary speed of the DSW. Fig.\ref{s_DSW_tunnel_trapping-fig} (a)-(b) shows the process of soliton tunneling through DSW. The initial value of the step-down makes the soliton amplitude larger. Fig.\ref{s_DSW_tunnel_trapping-fig} (c)-(d) shows soliton trapped in DSW.
\par
\subsection{For $x_0<0$}

Firstly, consider the case placing an initial soliton on the left side of the box, i.e., $x_0<0$.
For convenience, some notations should be introduced in advance. The amplitude of the soliton located in a certain region is represented by $a_{\Lambda}$, and the background wave is represented by $\overline{u}_{\Lambda}$, where the index $\Lambda \in \{R,L,\uppercase\expandafter{\romannumeral1}, \uppercase\expandafter{\romannumeral2}, \uppercase\expandafter{\romannumeral3} \}$. The Roman serial numbers have been marked in each area of Fig. \ref{lambda-whitham-fig}, where $R$ represents the plateau on the right and $L$ represents the plateau on the left.
\subsubsection{Soliton$-$mean flow transmission}
According to the previous preparations, the conditions for the interaction of soliton with RW and DSW have been given. Now the soliton starts moving from the left side with an initial speed of $2a_L$. As long as $a_L>2(\overline{u}_{\uppercase\expandafter{\romannumeral2}}-\overline{u}_L)=2U_0$ is satisfied, it will definitely be able to pass through the RW region and the DSW region. Approximately initialize the box initial value and modulated trial soliton by using Riemann invariants in the following form
\begin{equation}
\begin{aligned}
\label{initial soliton1}
u(x,0;x_0)=&\lambda_1-\lambda_2+\lambda_3\\
&+\left(2\lambda_{45}-2(\lambda_1-\lambda_2+\lambda_3)\right)
\mathrm{sech}^2(\sqrt{\lambda_{45}-(\lambda_1-\lambda_2+\lambda_3)}(x-x_0)),
\end{aligned}
\end{equation}
where $\lambda_1=0$ and
$$\lambda_2=\left\{
\begin{aligned}
&0, x<l\\
&U_0, x>l,
\end{aligned}
\right. \quad
\lambda_3=\left\{
\begin{aligned}
&0, x<0\\
&U_0, x>0.
\end{aligned}
\right. \quad
$$
\par
When considering soliton tunneling, take the moment when the interaction region appears as the critical time and consider two cases below.
\par
{\bf Case (a)}
\par
\begin{figure}
\centering
\includegraphics[width=6cm]{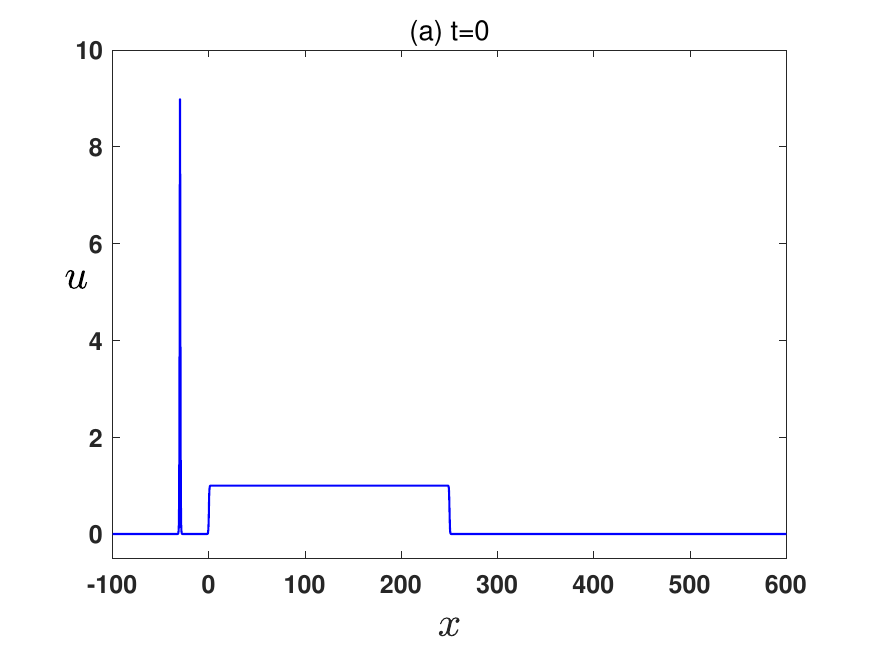}
\includegraphics[width=6cm]{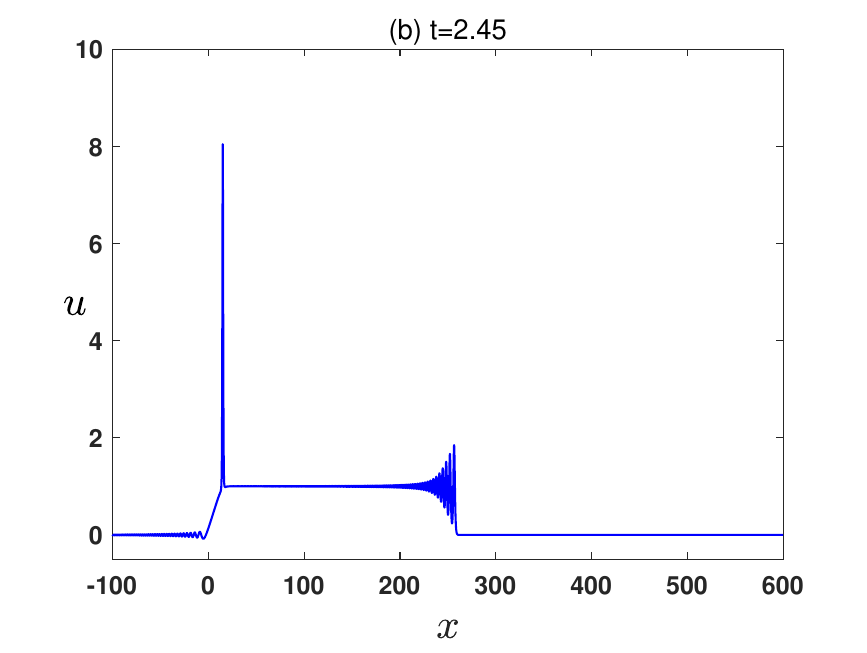}\\
\includegraphics[width=6cm]{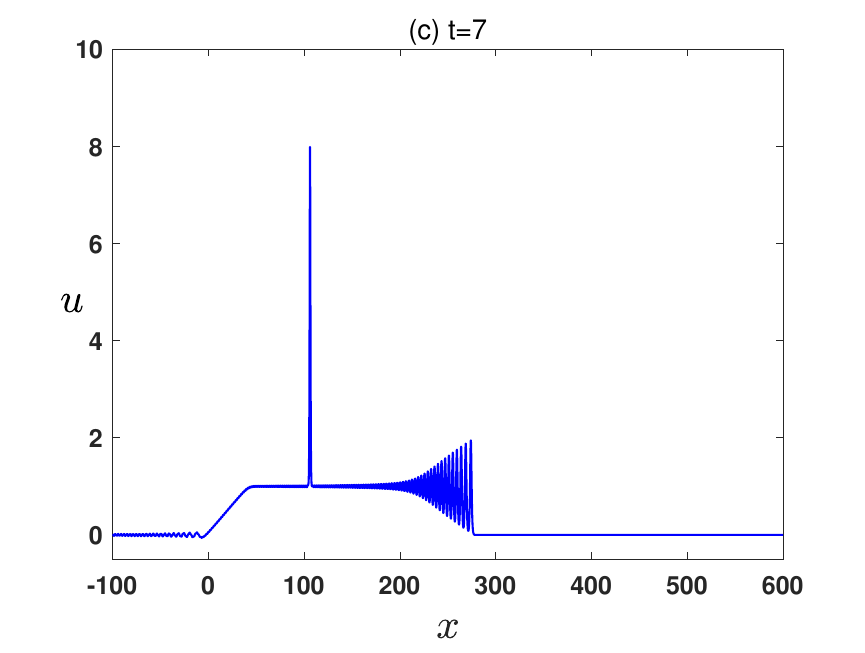}
\includegraphics[width=6cm]{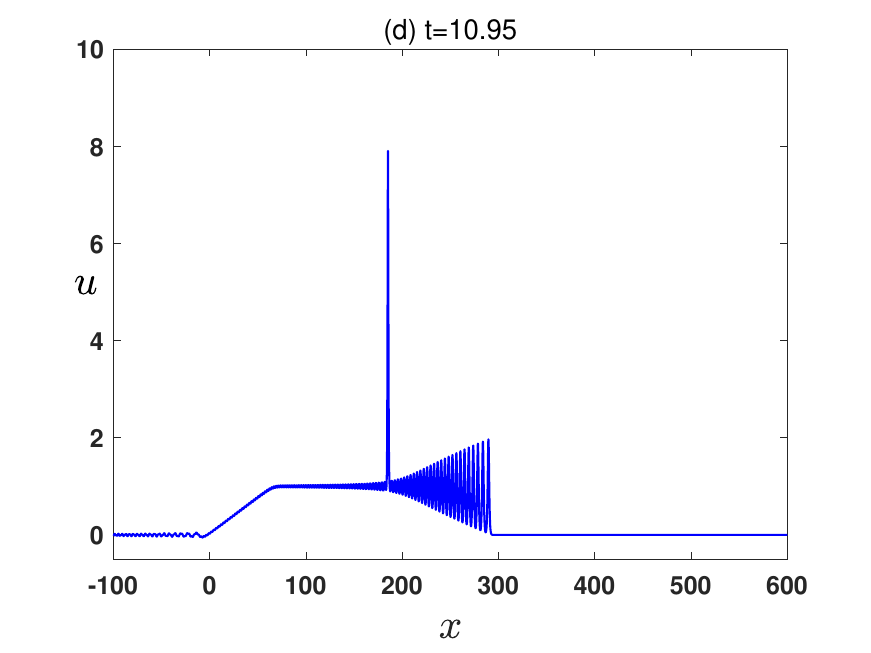}\\
\includegraphics[width=11.5cm]{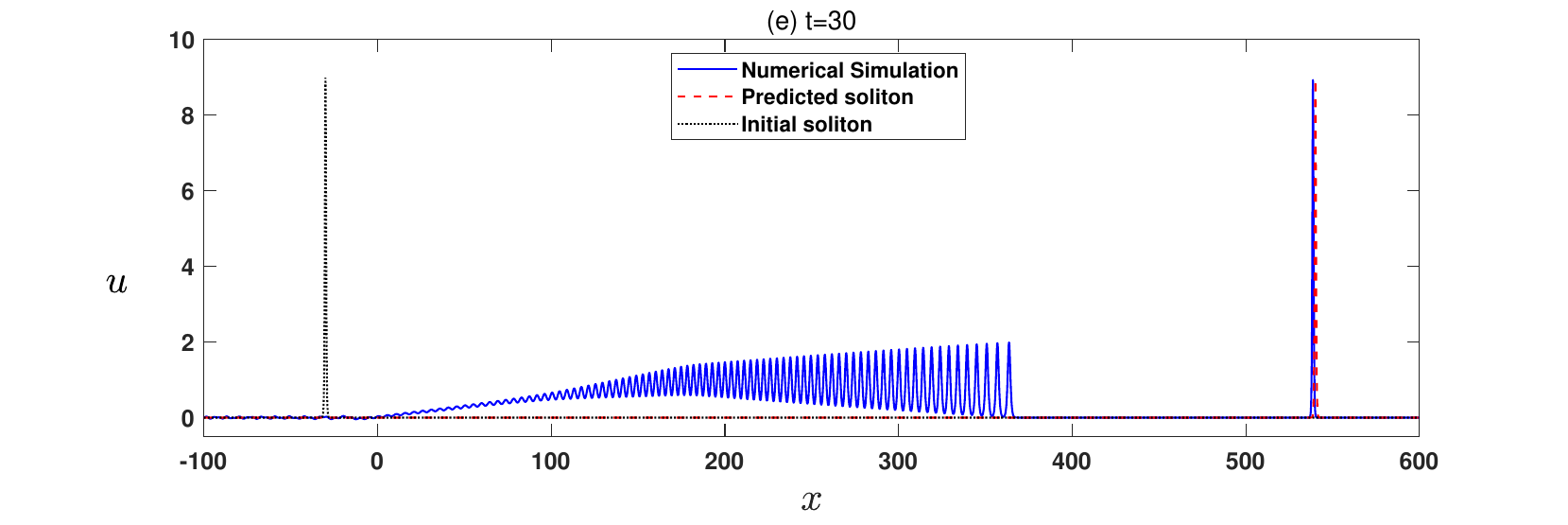}
\includegraphics[width=11.6cm]{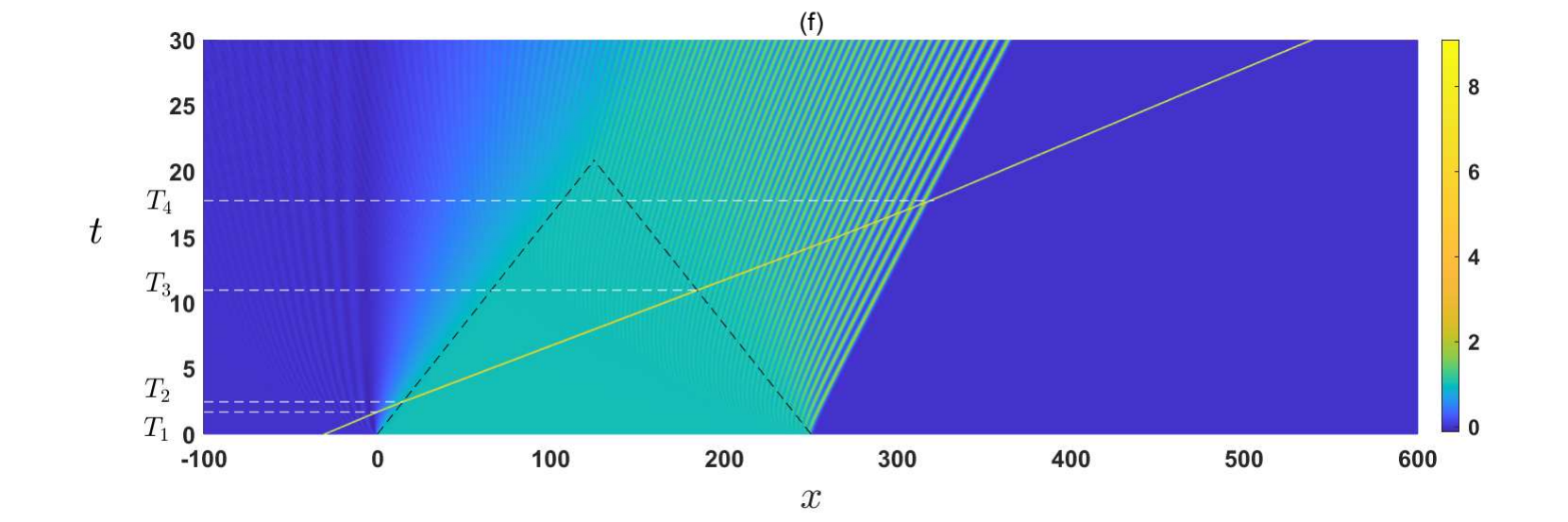}
\caption{{\protect\small Soliton tunneling occurs before the plateau disappears. (a) Box type initial value and initial trial soliton; (b) The trial soliton exactly leaves the RW area; (c) The trial soliton propagates to the right in the plateau region that has not yet disappeared; (d) The trial soliton enters the DSW region; (e) The trial soliton tunnels completely; (f) The process of soliton tunneling. The blue solid line shows the result of numerical simulation, the red dashed line is the result predicted by modulation theory, and the black dashed line displays the initial soliton. The initial condition is set as $U_0=1, l=250, x_0=-30, a_L=9$. The predicted amplitude is $a_R=9$ and the predicted phase shift $\delta$ is 0, while the numerical results are $a_R=8.928$ and $\delta=-0.121$.}}
\label{tunnel-duibi1-fig}
\end{figure}

If solitons tunneling  reaches before the plateau area disappears, it can be considered that this is a process of soliton passing through rarefaction wave and dispersive shock wave successively (see Fig. \ref{tunnel-duibi1-fig}).
\begin{figure}
\centering
\includegraphics[width=12.0cm]{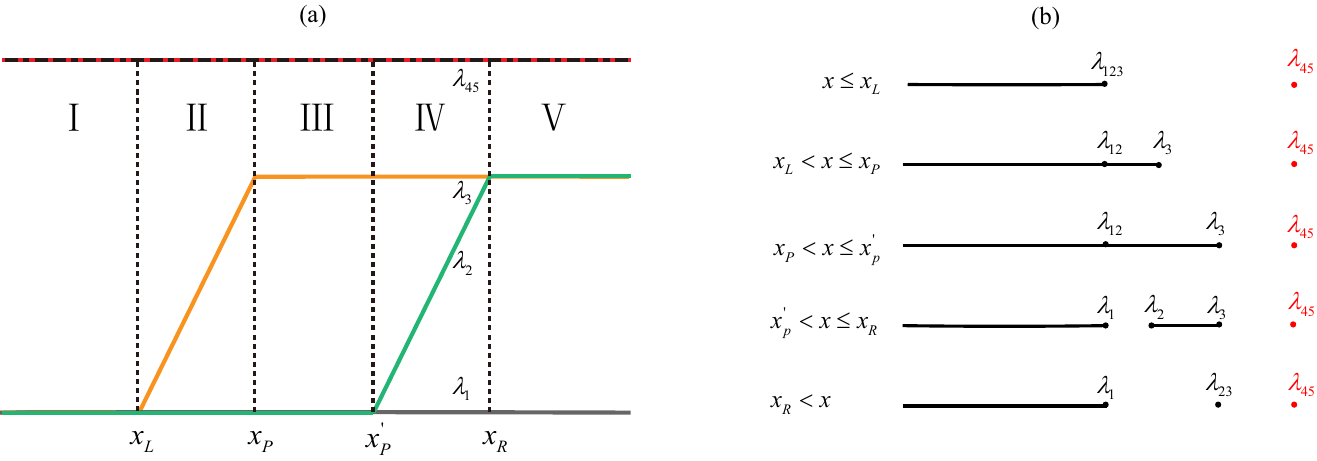}
\caption{{\protect\small Soliton tunneling occurs before the plateau disappears. (a) Evolutions of Riemann invariants; (b) The behaviors of the corresponding spectrums.}}
\label{tunnel-whitham1-fig}
\end{figure}
The RW solution can be given explicitly
\begin{equation}
\overline{u}(x,t)=\left\{
\begin{aligned}
&0,  &x\leq 0,\\
&\frac{x}{6t},  &0<x<6U_0t,\\
&U_0,  &6U_0t<x,
\end{aligned}
\right.
\label{RW-solution}
\end{equation}
and the DSW is modulated by the Whitham equation (\ref{KdV-Whitham-E}) and satisfies
$$
\lambda_1=0, \quad \lambda_3=U_0, \quad v_2(0,\lambda_2,U_0)=\frac{x}{t}.
$$
Because the interaction region has not yet appeared, the boundaries (\ref{before-bound}) can be calculated directly from the Whitham equation.
Now we hope to give the parameter conditions for solitons to complete tunneling before the plateau disappears. The moments when the soliton enters and exits each region are marked in Fig. \ref{tunnel-duibi1-fig}(f). From the initial position to the left boundary of the RW region (i.e., $x=0$), the soliton propagates at a speed of $2a_L$ during this process. Directly, the trial soliton enters the RW region at time $T_1=-\frac{x_0}{2a_L}$. In fact, according to the RW solution (\ref{RW-solution}), the soliton amplitude can be determined by (\ref{transmission}) as
\begin{equation}
a(x,t)=a_L-2\overline{u}(x,t)=\left\{
\begin{aligned}
&a_L,  &x\leq 0,\\
&a_L-\frac{x}{3t},  &0<x<6U_0t,\\
&a_{\uppercase\expandafter{\romannumeral2}},  &6U_0t<x.
\end{aligned}
\right.
\end{equation}
The trajectory of the soliton can be determined by a Cauchy problem
\begin{equation}
\left\{
\begin{aligned}
&\frac{d x_s}{dt}=C,\\
&x_s(0)=x_0,
\end{aligned}
\right.
\end{equation}
where $C$ is the speed of the soliton. Then, the moment $T_2$ when the trial soliton leaves the RW region can be obtained below
\begin{equation}
\label{T2}
T_2=-(\frac{a_L}{a_{\uppercase\expandafter{\romannumeral2}}})^{\frac{3}{2}}\frac{x_0}{2a_L}.
\end{equation}
On plateau \uppercase\expandafter{\romannumeral2}, the soliton maintains a speed of $6U_0+2a_{\uppercase\expandafter{\romannumeral2}}$ and propagates to the right. Then the moment and position of entering the DSW region is $(T_3,x_3)$, and the boundary of the DSW determined by the Whitham velocity can determine $T_3$ as
\begin{equation}
\label{T3}
T_3=\frac{l+2a_{\uppercase\expandafter{\romannumeral2}}T_2}{12U_0+2a_{\uppercase\expandafter{\romannumeral2}}},
\end{equation}
where $a_{\uppercase\expandafter{\romannumeral2}}=a_L-2U_0$.
\par
The soliton trajectory in the DSW region is more complicated because the mean flow is a modulated Jacobian elliptic cn function. According to the equations (\ref{q}) and (\ref{C}), it follows that
\begin{equation}
\left\{
\begin{aligned}
&\frac{d x_s}{dt}=C=q+2\overline{u}_{\uppercase\expandafter{\romannumeral2}}=2a_L+2\overline{u}_{\uppercase\expandafter{\romannumeral2}},\\
&x_s^{initial}=x_3.
\end{aligned}
\right.
\end{equation}
Approximately, we might as well use the middle value of the fluctuation range $U_0$ to replace the original modulated wave $\overline{u}_{\uppercase\expandafter{\romannumeral2}}$, and then integrate it to get
\begin{equation}
\label{T4}
T_4=\frac{l+2a_{\uppercase\expandafter{\romannumeral2}}T_2}{2a_L-2U_0}.
\end{equation}
The moments $T_1, T_2, T_3$ and $T_4$ have been marked in Fig. \ref{tunnel-duibi1-fig}(f). Lastly, it should be ensured that the parameter conditions satisfy $T_4<\frac{l}{12U_0}$, so that the soliton will complete tunneling before the plateau disappears.
\par
In what follows, this process will be described within the framework of Whitham modulation theory. For the Riemann problem of the KdV equation with $u_L>u_R$, an oscillating dispersive shock wave is always generated, or it can also be considered that a gradient mutation occurs. The method of regularization here is to increase the phase of the modulation. Following the procedure of \cite{Luo-stud-2023},
we construct Riemann invariants and use the two-genus Whitham equation to describe this process.
For the soliton tunneling problem we are concerned about, the limit state of Whitham velocities are indispensable. More precisely, the interaction of soliton with the mean flow relies on the limit state $(\lambda_4\rightarrow \lambda_5)$ of the two-genus Whitham modulation system to be described. After some calculations, it is found that when $k\in{\{1,2,3\}}$, the Whitham velocities in two-genus Whitham equation can degenerate into the Whitham velocities of one-genus at the limit state $\lambda_4\rightarrow \lambda_5$, which are
\begin{equation}
\begin{aligned}
{\rm lim}_{\lambda_4\rightarrow\lambda_5}v^{(2)}_1&=v_1,\quad
{\rm lim}_{\lambda_4\rightarrow\lambda_5}v^{(2)}_2=v_2,\quad
{\rm lim}_{\lambda_4\rightarrow\lambda_5}v^{(2)}_3=v_3,\\
{\rm lim}_{\lambda_4\rightarrow\lambda_5}v^{(2)}_4&={\rm lim}_{\lambda_4\rightarrow\lambda_5}v^{(2)}_5 \equiv v_{45},\\
&=2(\lambda_1+\lambda_2+\lambda_3)+\frac{4(\lambda_{45}-\lambda_2)}
{1-\sqrt{(\frac{(\lambda_{45}-\lambda_2)(\lambda_{3}-\lambda_1)}{(\lambda_{45}-\lambda_3)(\lambda_{45}-\lambda_1)})}}\mathrm{Z}(\psi,m),
\end{aligned}
\end{equation}
where $\mathrm{sin} \psi=\sqrt{\frac{\lambda_{45}-\lambda_3}{\lambda_{45}-\lambda_2}}$ and $\mathrm{Z}(\cdot,\cdot)$ is the Jacobian zeta function. Note that ${\rm lim}_{\lambda_1\rightarrow \lambda_2}({\rm lim}_{\lambda_4\rightarrow \lambda_5} v_3^{(2)})={\rm lim}_{\lambda_1\rightarrow \lambda_2}v_3=6\lambda_3$. That is to say, when $\lambda_1\rightarrow \lambda_2$ and $\lambda_4\rightarrow \lambda_5$, regions \uppercase\expandafter{\romannumeral1}, \uppercase\expandafter{\romannumeral2}, and \uppercase\expandafter{\romannumeral3} degenerate into the Hopf equation satisfied by $\lambda_3$.
The Whitham equations here are
\begin{equation}
\label{whitham123}
\begin{aligned}
&\frac{\partial \lambda_{12}}{\partial t}+(12\lambda_{12}-6\lambda_{3})\frac{\partial \lambda_{12}}{\partial x}=0,\\
&\frac{\partial \lambda_{3}}{\partial t}+6\lambda_{3}\frac{\partial \lambda_{3}}{\partial x}=0,\\
&\frac{\partial \lambda_{45}}{\partial t}+(2\lambda_{3}+4\lambda_{45})\frac{\partial \lambda_{45}}{\partial x}=0,
\end{aligned}
\end{equation}
where $\lambda_{45}$ is a global invariant.
\par
Construct Riemann invariant $\lambda_{45}$ by
\begin{equation}
\label{lambda45}
\lambda_{45}=\overline{u}+\frac{1}{2}a,
\end{equation}
which is the bridge between solitonic modulation system and Whitham modulation equation. Substituting (\ref{lambda45}) into (\ref{whitham123}), and reminding $\overline{u}=\lambda_{3}$, this can exactly match the system (\ref{modulation}).
\par
The trajectory of the trial soliton in regions \uppercase\expandafter{\romannumeral1}, \uppercase\expandafter{\romannumeral2} and \uppercase\expandafter{\romannumeral3}  can be described by the following equation
\begin{equation}
\left\{
\label{s-whitham123}
\begin{aligned}
&\frac{d x_s}{dt}=v_{45}(0,0,\lambda_3(\frac{x_s(t)}{t}), \lambda_{45}),\\
&x_s^{initial}=x_0,
\end{aligned}
\right.
\end{equation}
while in regions \uppercase\expandafter{\romannumeral3}, \uppercase\expandafter{\romannumeral4} and \uppercase\expandafter{\romannumeral5}, it is described by
\begin{equation}
\left\{
\label{s-whitham345}
\begin{aligned}
&\frac{d x_s}{dt}=v_{45}(0,\lambda_2(\frac{x_s(t)}{t}), U_0, \lambda_{45}),\\
&x_s^{initial}=x_3.
\end{aligned}
\right.
\end{equation}
In the connected region \uppercase\expandafter{\romannumeral3}, the matching of (\ref{s-whitham123}) and (\ref{s-whitham345}) can be checked according to the limit of the Whitham velocities. Due to the global invariance $\lambda_{45}$, it can be seen that the amplitude of the soliton does not change before and after tunneling, i.e.,
$$
\begin{aligned}
&\overline{u}_L+\frac{1}{2}a_L=\overline{u}_{\uppercase\expandafter{\romannumeral2}}+\frac{1}{2}a_{\uppercase\expandafter{\romannumeral2}}=\overline{u}_R+\frac{1}{2}a_R,\\
&a_L=a_R.
\end{aligned}
$$
The change in phase can be calculated according to the equation (\ref{invariance}), that is
\begin{equation}
\frac{k_L}{k_R}=1.
\end{equation}
\par
According to the conservation of wave number, it can be seen that the phase does not change, which is consistent with the conclusion obtained by the transmission condition (\ref{transmission}) and phase condition (\ref{invariance}). Fig. \ref{tunnel-duibi1-fig} shows the process of a trial soliton crossing the box type barrier before disappearing from the plateau, and the critical moments $T_1=-\frac{x_0}{2a_L}$, $T_2$ in (\ref{T2}), $T_3$ in (\ref{T3}) and $T_4$ in (\ref{T4}) given by the soliton velocity and trajectory equation are numerically checked. Moreover, the theoretically predicted soliton amplitude and phase are in good agreement with the numerical results (see Fig. \ref{tunnel-duibi1-fig}(e) for details).
\par
Fig. \ref{tunnel-whitham1-fig} shows the evolutions of the Riemann invariant during the process of soliton tunneling and the changes of spectrums when the trial soliton tunnels before disappearing from the plateau. Regions \uppercase\expandafter{\romannumeral1}, \uppercase\expandafter{\romannumeral2} and \uppercase\expandafter{\romannumeral3} appear as shrinked spectral point and a semi-infinite spectral band, which corresponds to solitons propagating on RW as background wave. Observe regions \uppercase\expandafter{\romannumeral3}, \uppercase\expandafter{\romannumeral4}, and \uppercase\expandafter{\romannumeral5} and notice that this is the behavior of the spectral band for the one-phase DSW with an additional spectral point representing the contraction of the soliton solution.
\par
{\bf Case (b)}
\par
\begin{figure}
\centering
\includegraphics[width=12.0cm]{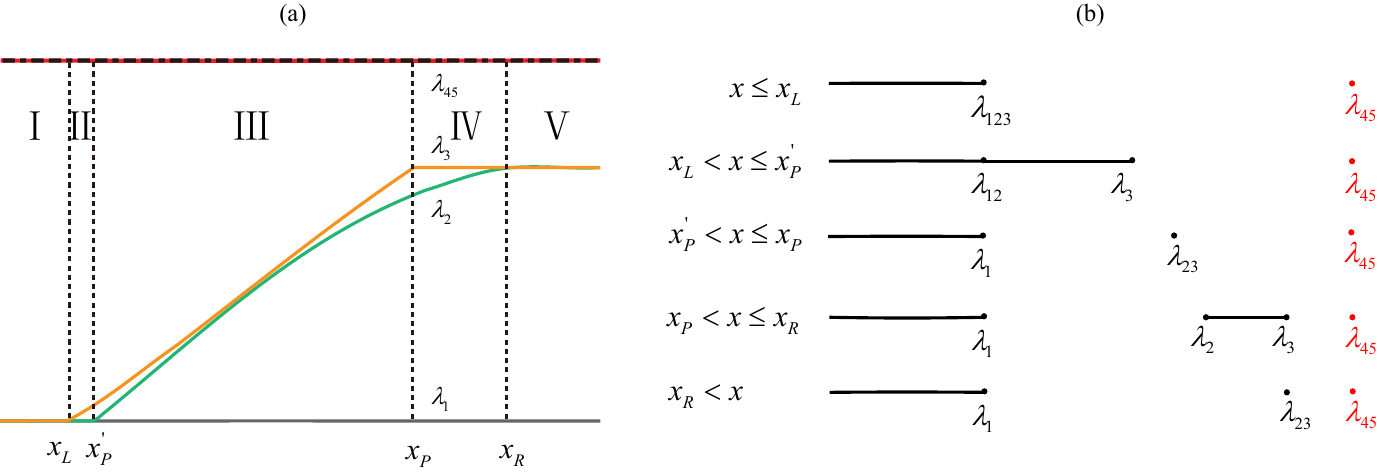}
\caption{{\protect\small Soliton tunneling occurs after the plateau disappears. (a) Evolutions of Riemann invariants; (b) The behaviors of the corresponding spectrums.}}
\label{tunnel-whitham2-fig}
\end{figure}
\begin{figure}
\centering
\includegraphics[width=6.0cm]{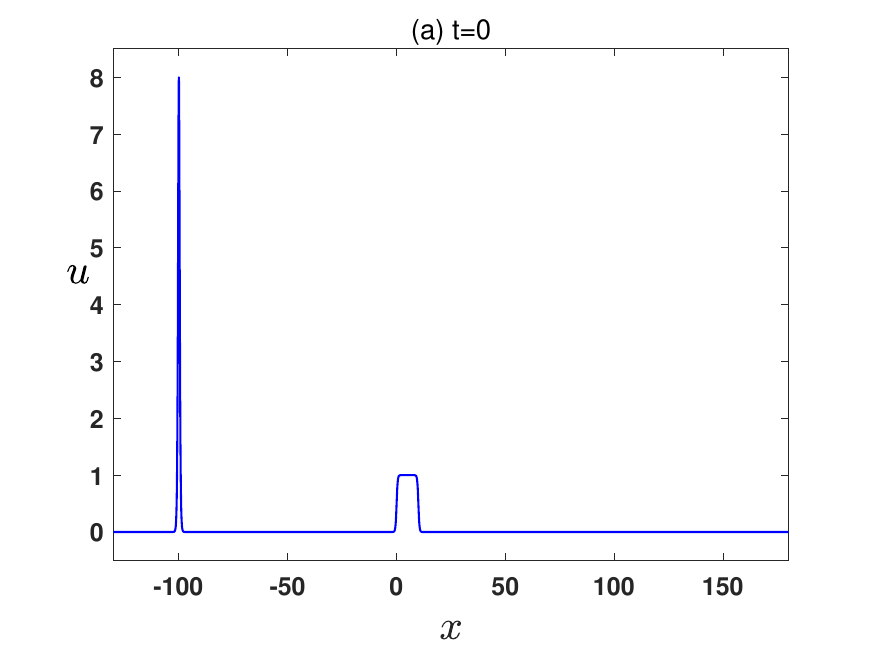}
\includegraphics[width=6.0cm]{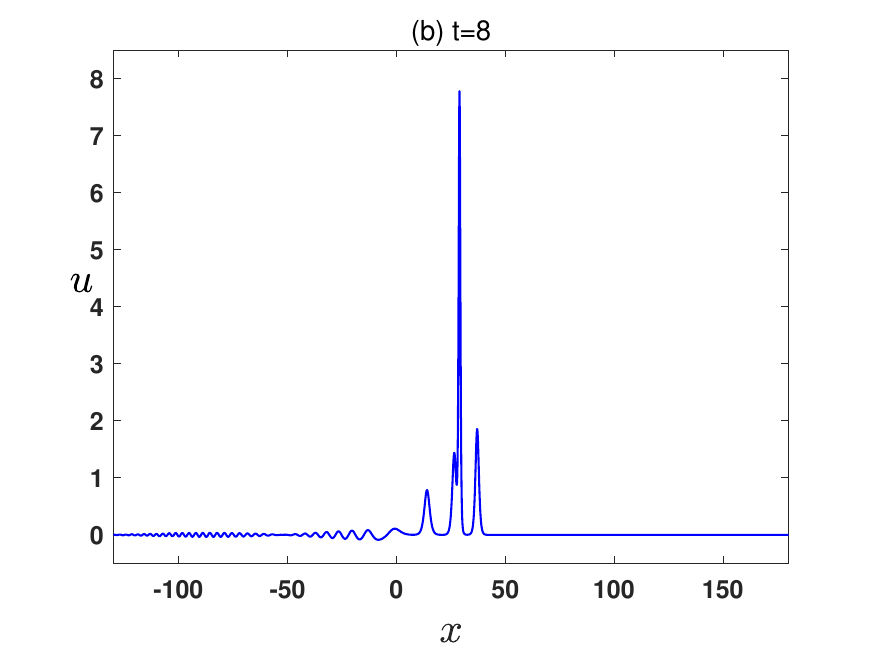}\\
\includegraphics[width=11.5cm]{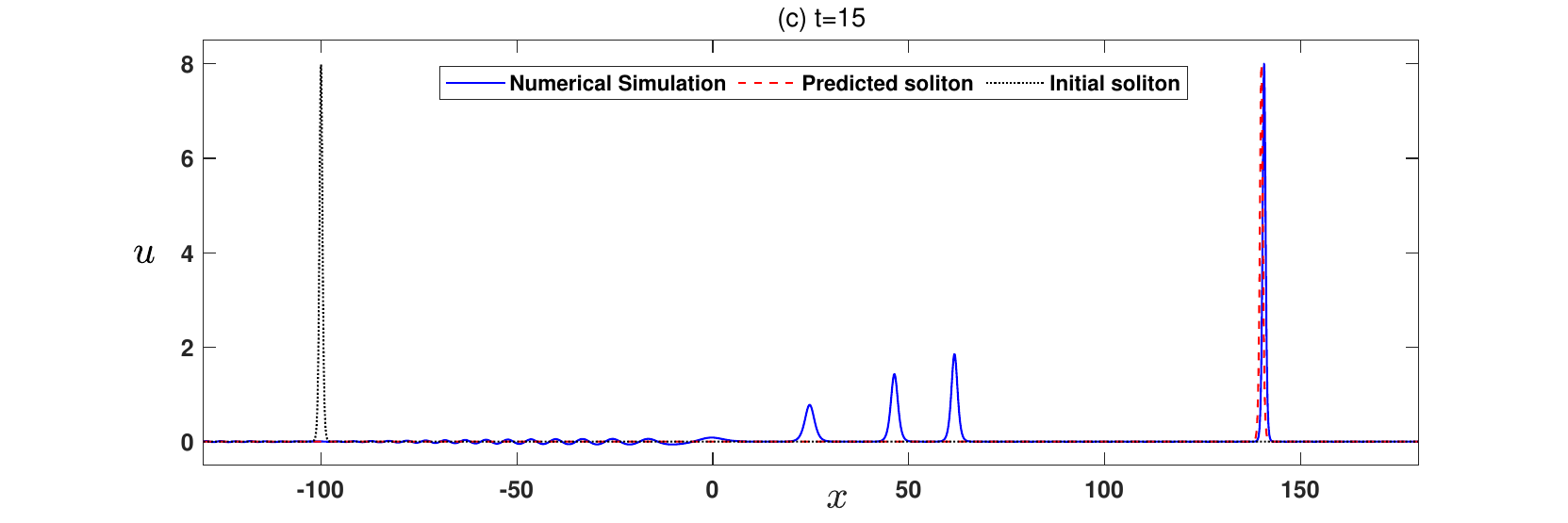}\\
\includegraphics[width=11.6cm]{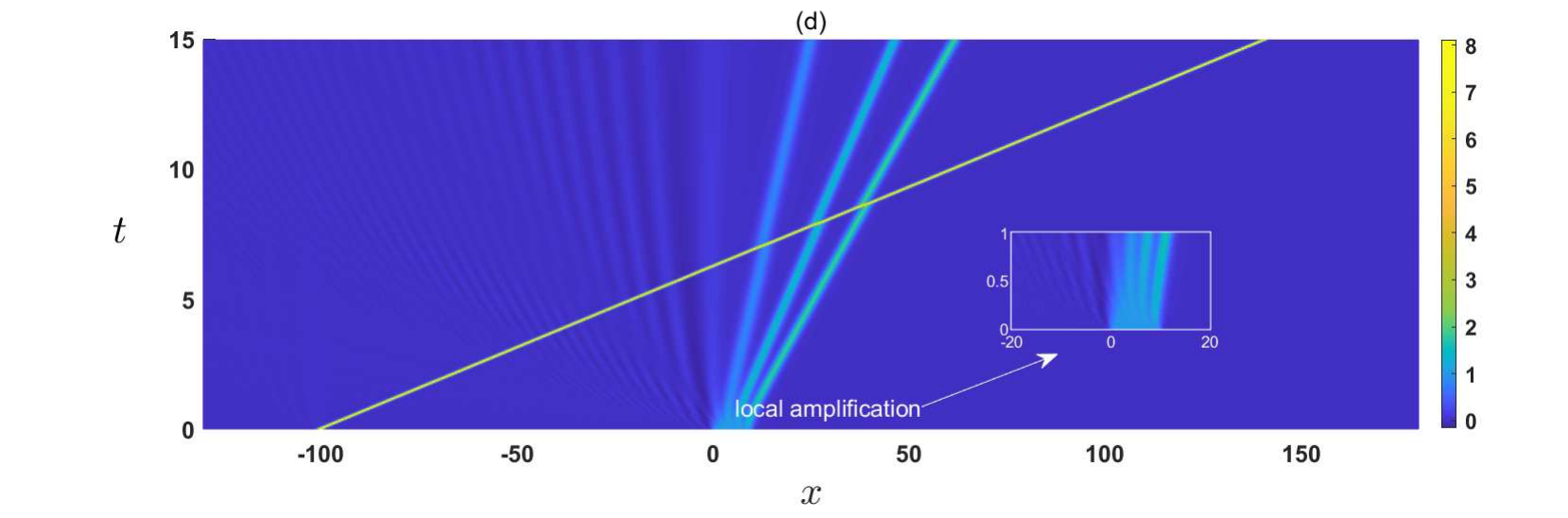}\\
\caption{{\protect\small Soliton tunneling occurs after the plateau disappears. (a) Box type initial value and initial trial soliton; (b) The trial soliton interacts with the soliton train at time $t=8$; (c) Soliton tunnels completely; (d) The detailed process of soliton tunneling in $(x,t)$ plane. The blue solid line shows the result of numerical simulation, the red dashed line is the result predicted by modulation theory, and the black dashed line displays the initial soliton. The initial condition is set to be $U_0=1, l=10, x_0=-100, a_L=8$. The predicted amplitude is $a_R=8$ and the predicted phase shift is $\delta=0$. Furthermore, the numerical results are $a_R=8.002$ and $\delta=0.013$.}}
\label{tunnel-duibi2-fig}
\end{figure}
In this section, we consider the case that soliton tunneling occurs after the interaction region appears. The distribution of Riemann invariants in this case is shown in Fig. \ref{tunnel-whitham2-fig}. The difference with Case (a) lies in the selection of $\lambda_{45}$, which determines the speed of the soliton. In fact, the occurrence of this case also depends on the distance $l$ between the two discontinuities. The process of soliton tunneling is to propagate rightward on the plateau into the RW region, interact with the soliton train after leaving the RW region, and finally enter the DSW region and pass out to reach the rightmost plateau. Fig. \ref{tunnel-duibi2-fig} shows the process of soliton tunneling through the soliton train. The theoretically predicted soliton phase and amplitude have acceptable errors compared with the numerical simulation results.
\par
In the framework of the KdV-Whitham modulation system, the soliton velocity in the plateau region on the left behaves as the limit state below and propagates on the background of $\overline{u}=\lambda_3$, that is
\begin{equation}
{\rm lim}_{\lambda_2\rightarrow \lambda_1}v_{45}=2\lambda_3+4\lambda_{45}.
\end{equation}
On the right plateau, the soliton velocity is
\begin{equation}
{\rm lim}_{\lambda_2\rightarrow \lambda_3}v_{45}=2\lambda_1+4\lambda_{45},
\end{equation}
where the background is $\overline{u}=\lambda_1$.
\par
Fig. \ref{tunnel-whitham2-fig}(b) shows the evolution of the spectrum for $x$ in different intervals. In region \uppercase\expandafter{\romannumeral3}, the spectrum band shrinks approximately to a point, and the value of the spectrum point becomes larger as the displacement becomes larger. In the diagram of the solution to the KdV equation in Fig. \ref{tunnel-duibi2-fig}, this appears as a train of solitons of increasing amplitude.
\par
The description of the trial soliton trajectory is similar to that of Case (a), but region \uppercase\expandafter{\romannumeral3} is different, that is
\begin{equation}
\left\{
\begin{aligned}
&\frac{d x_s}{dt}=v_{45}(0,\lambda_2(\frac{x_s(t)}{t}),\lambda_3(\frac{x_s(t)}{t}), \lambda_{45}),\\
&x_s^{initial}=x_P^{'}.
\end{aligned}
\right.
\end{equation}
The speed of the trial soliton is expressed as $v_{45}$. The strict hyperbolicity of the KdV equation ensures that $v_{45}>v_{3}>v_{2}>v_{1}$, which indicates that the soliton always tunnels.
In other words, according to the equation (\ref{initial soliton1}), the speed of the modulated soliton train is $v=2\lambda_1+4\lambda_{23}$, and the order relationship of the Riemann invariant ensures that the speed of the trial soliton is always greater than the speed of any soliton in the soliton train.

\subsubsection{Soliton$-$mean flow trapping}
\begin{figure}
\centering
\includegraphics[width=11.0cm]{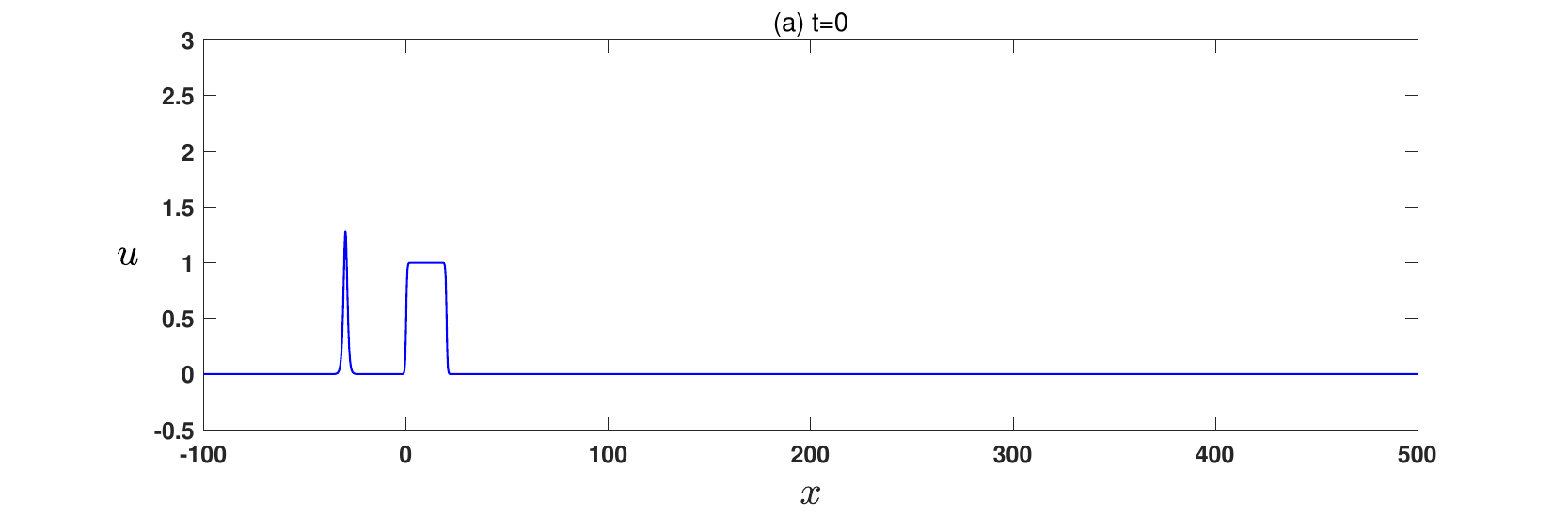}
\includegraphics[width=11.0cm]{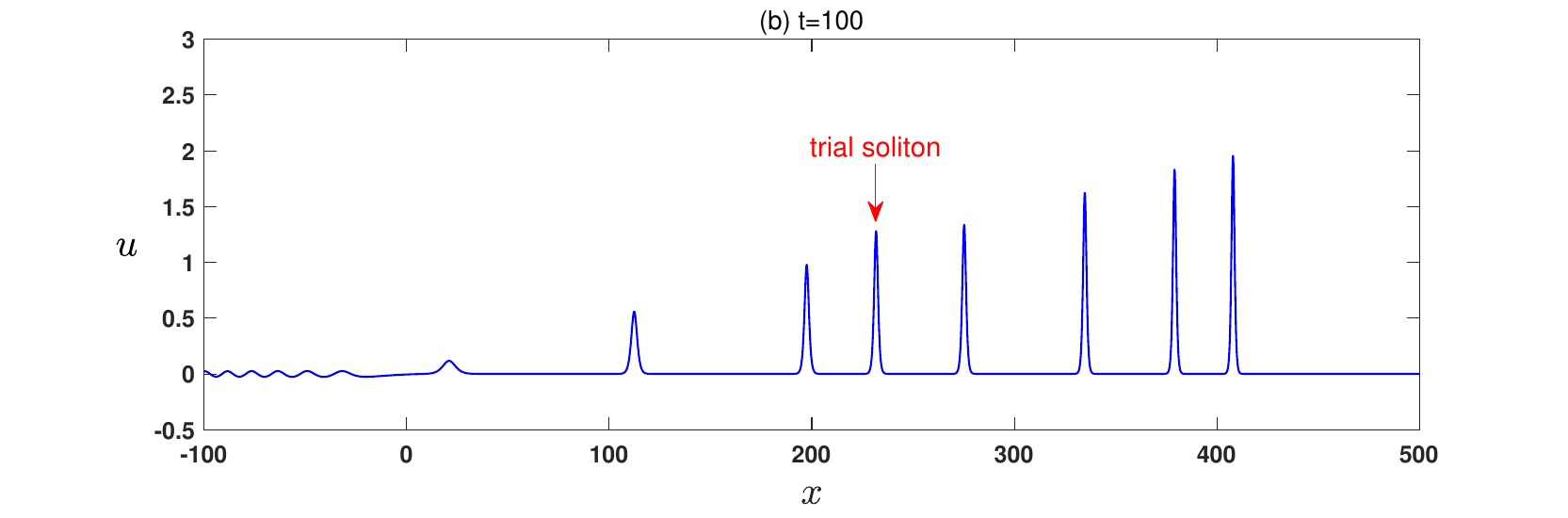}
\includegraphics[width=11.1cm]{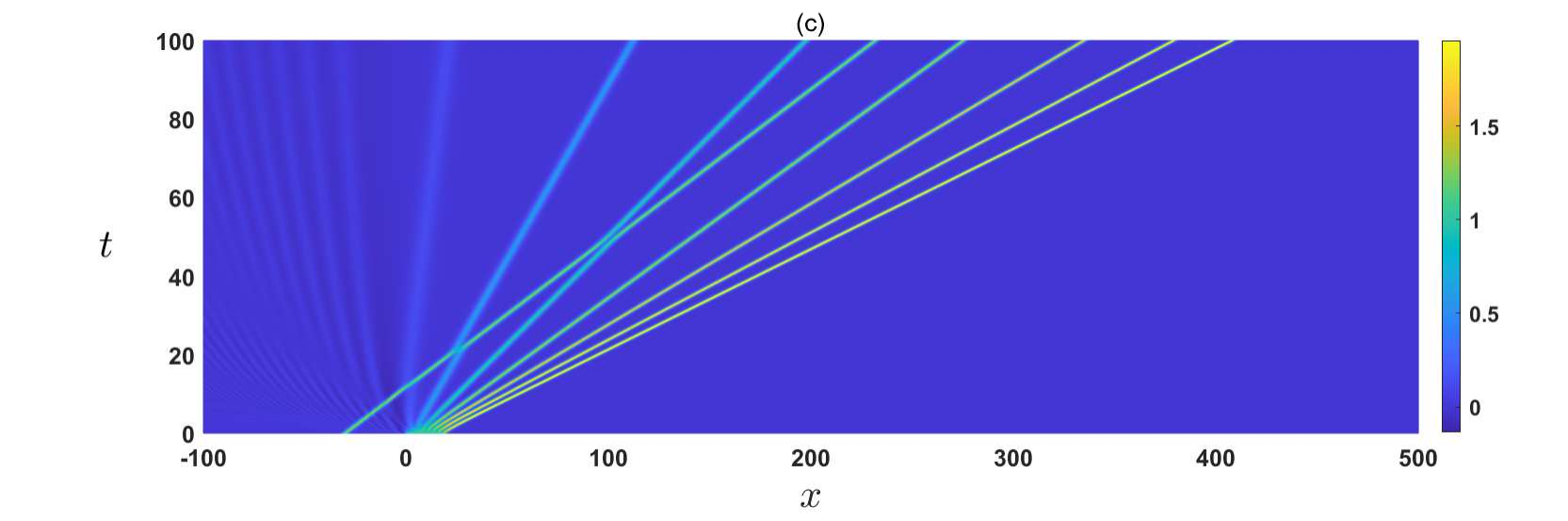}
\caption{{\protect\small The trial soliton is trapped in the interaction region. (a) Box type initial value and initial trial soliton; (b) The trial soliton is trapped completely; (c) The process of soliton trapping. The initial condition is set to be $U_0=1, l=20, x_0=-30, a_L=1.28$.}}
\label{trapping-duibi-fig}
\end{figure}
\begin{figure}
\centering
\includegraphics[width=12.0cm]{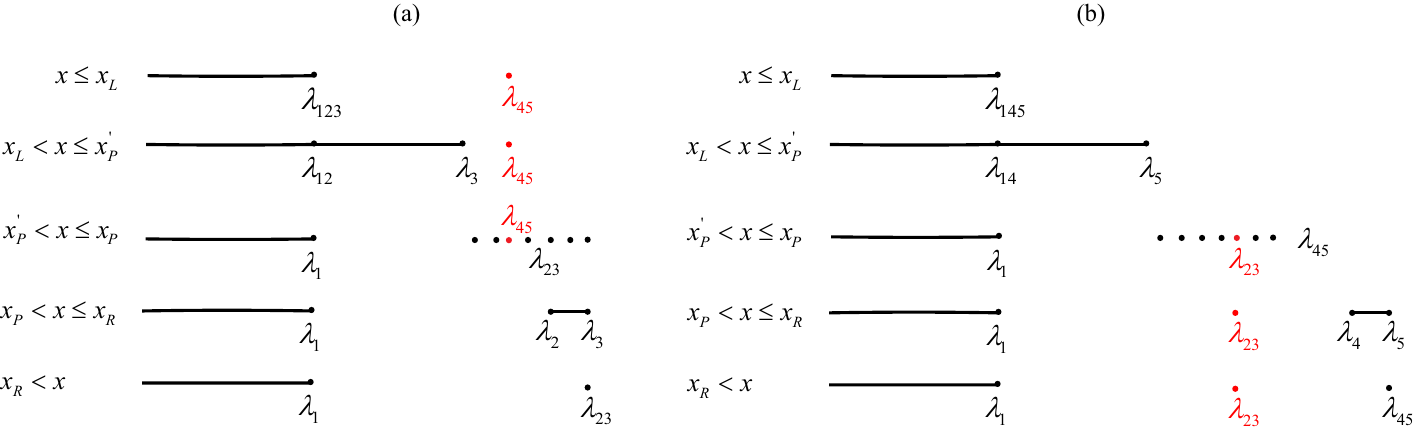}
\caption{{\protect\small The trial soliton is trapped in the interaction region. (a) The behavior of the corresponding spectrums when $x_0<0$; (b) The behavior of the corresponding spectrum when $x_0>l$.}}
\label{trappingl2-whitham-fig}
\end{figure}
According to the long-term evolutions of the initial value of barrier type (\ref{Initial condition}), there are five regions in the $(x,t)$ plane at each time: the plateau region, the RW region, the interaction zone (or the plateau region), the DSW region and the plateau region. Still when $x_0<0$, that is, the initial position of the soliton is on the left, the soliton trapping will be discussed. It is clarified that the condition for soliton trapping is $a_L<2U_0$. According to (\ref{BC-IZ-left}), it can be seen that when time $t\rightarrow\infty$, that is, the time is long enough, the two boundaries of the RW region gradually approach. Therefore, in the long-term sense, there is no soliton trapped in the RW region. In fact, if the soliton starts from the left, it will definitely tunnel through the DSW instead of being trapped, which is already clear during the soliton-DSW interaction. So what we focus on is the case where soliton is trapped in the interaction region.
\par
After a period of time, the potential barrier evolves into a soliton train that almost covers the wave region. The trial soliton elastically collides with the soliton train from the left until the speed is between two solitons in the soliton train (see Fig.\ref{trapping-duibi-fig}(b)). In order to demonstrate the phenomenon of soliton trapping, Fig. \ref{trapping-duibi-fig} is provided to show the initial state and the evolutions of soliton trapping in details.
\par
Fig. \ref{trappingl2-whitham-fig}(a) shows the distribution of Riemann invariants of solitons trapped in the interaction region and the varying behavior of the spectrum. We construct $\lambda_{45}$ as the spectral point corresponding to the trial soliton. In the interval $x_P^{'}<x\leq x_P$, the spectral band shrinks to spectral points. This column of varying spectral points ($\lambda_{2}$ and $\lambda_{3}$ are varying) appears as a soliton sequence, and $x>x_R$ corresponds to $\lambda_{2 }=\lambda_{3}=U_0$ (constant), what is generated is the soliton front. Using $x_*$ to mark the criticality of soliton trapping in the soliton column, then when $x<x_*$, the soliton speed can be expressed by
\begin{equation}
\begin{aligned}
v_{45}
&=2(\lambda_1+\lambda_2+\lambda_3)+\frac{4(\lambda_{45}-\lambda_2)}{1-\sqrt{\frac{(\lambda_2-\lambda_{45})(\lambda_1-\lambda_3)}{(\lambda_1-\lambda_{45})(\lambda_3-\lambda_{45})}}\mathrm{Z}(\psi_{45},m_{45})},
\end{aligned}
\end{equation}
where $\mathrm{sin}\psi_{45}=\sqrt{\frac{\lambda_3-\lambda_{45}}{\lambda_2-\lambda_{45}}}$, and $m_{45}=\frac{\lambda_2-\lambda_1}{\lambda_3-\lambda_1}$.
When the trial soliton is trapped in the soliton train, the trial soliton will maintain a speed of $2\lambda_1+4\lambda_{45}$ and propagate to the right.
\par
To explain the trapping behavior in Fig.\ref{trapping-duibi-fig}(b)-(c), the trial soliton trapped here will not overlap with one of the soliton train. This is related to the fact that there are no double zeros in the KdV equation spectral problem, i.e., the linear Schr\"{o}dinger equation. Therefore, the trial soliton is always trapped between two solitons in the soliton train, as shown in Fig. \ref{trapping-duibi-fig}.
\subsection{For $0<x_0<l$}
\begin{figure}
\centering
\includegraphics[width=11.0cm]{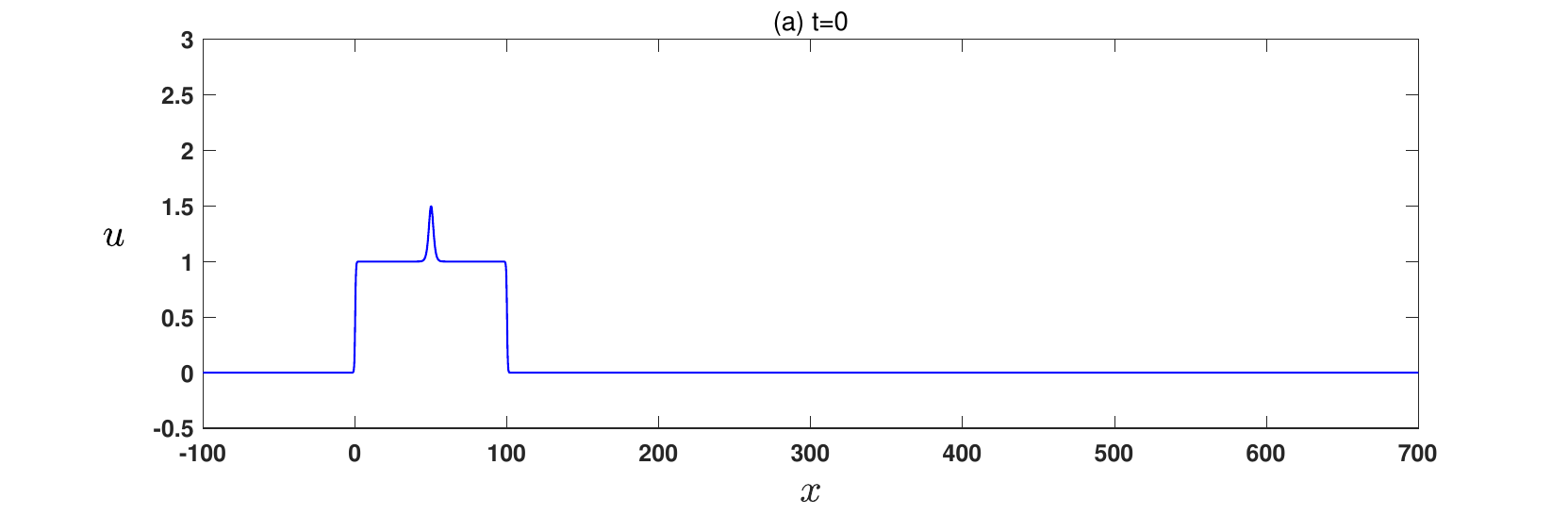}
\includegraphics[width=11.0cm]{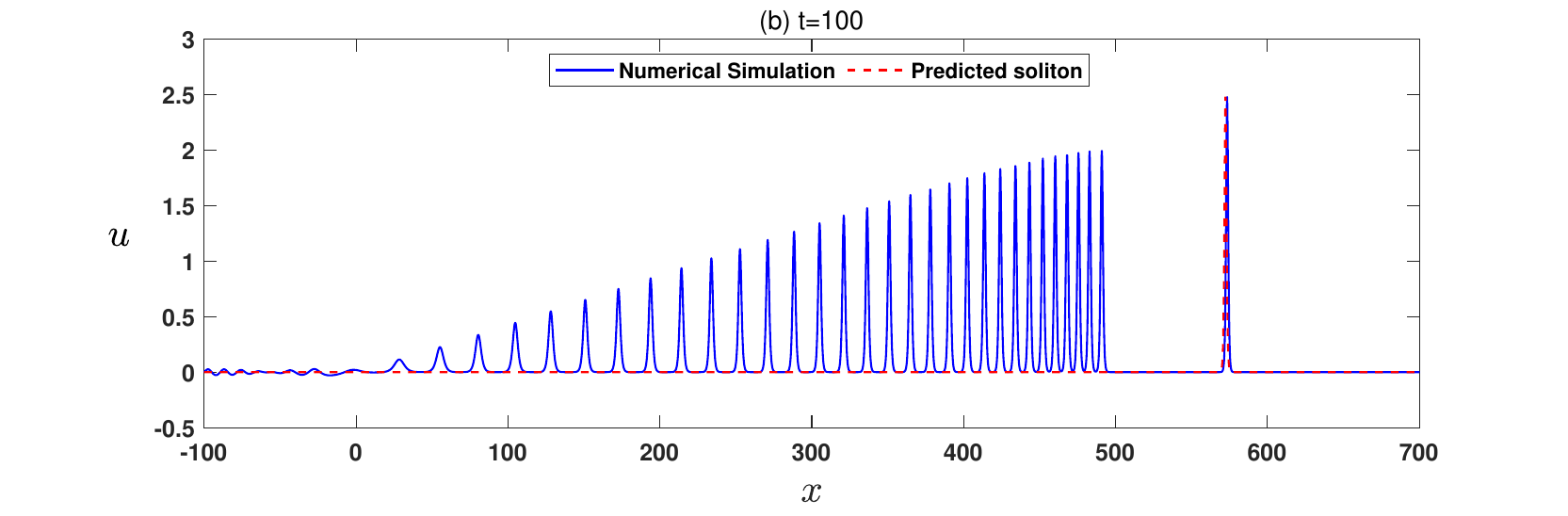}
\includegraphics[width=11.1cm]{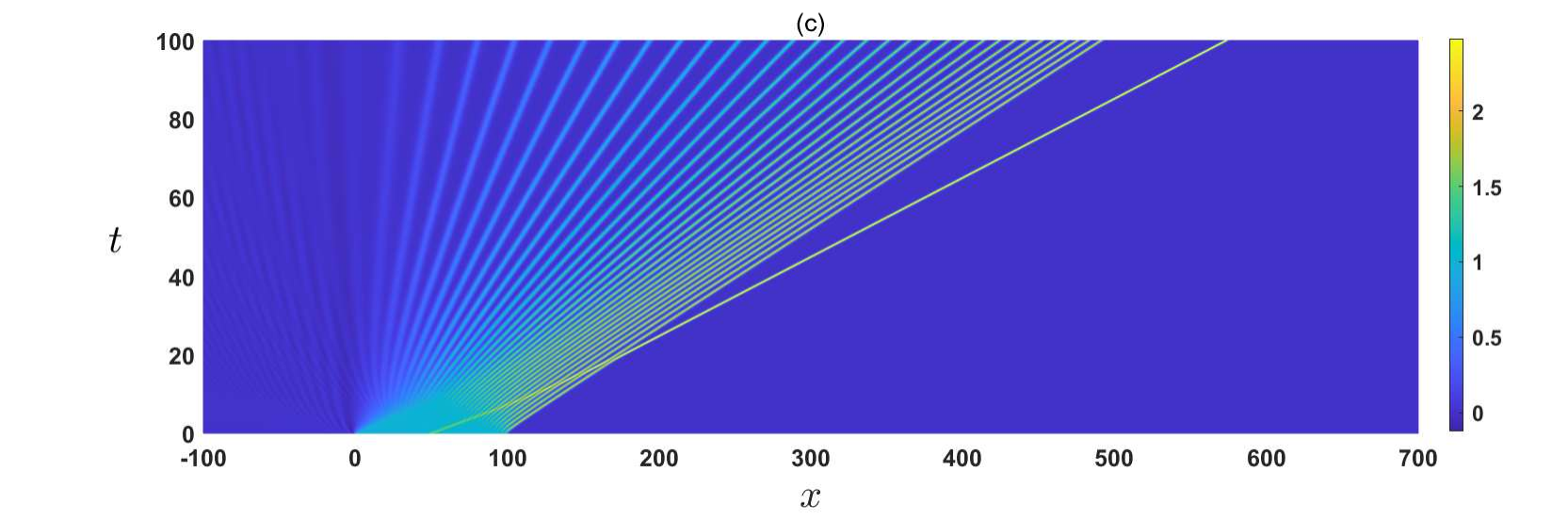}
\caption{{\protect\small Soliton tunneling when $0<x_0<l$. (a) Box type initial value and initial trial soliton; (b) Soliton tunnels completely; (c) The process of soliton tunneling. The blue solid line shows the result of numerical simulation, and the red dashed line displays the result predicted by modulation theory. The initial condition is set to be $U_0=1, l=100, x_0=50, a_L=0.5$. The predicted amplitude is $a_R=2.5$ and the predicted phase shift is $\delta=27.6393$, and fortunately the numerical results are $a_R=2.4787$ and $\delta=27.5435$.}}
\label{middle-duibi-fig}
\end{figure}
If the initial soliton is located in the middle of two discontinuities (scale separation is still guaranteed), the soliton propagating to the right will not interact with the RW, and it is asserted that the soliton must tunnel rather than be trapped in a certain region. If the distance between the two discontinuities is very large, we are not interested in the case that the soliton is in the middle, because this is almost similar to the interaction of the soliton with the DSW produced by a single discontinuity on the right \cite{Luo-stud-2023}. In fact, such placement of the initial trial soliton only results in soliton tunneling (see Fig. \ref{middle-duibi-fig}). An assertion will be illustrated below.
\par
Approximately initialize the box initial value and modulated soliton by using Riemann invariants in the following form
$$
\begin{aligned}
u(x,0;x_0)=&\lambda_1-\lambda_2+\lambda_3\\
&+\left(2\lambda_{45}-2(\lambda_1-\lambda_2+\lambda_3)\right)\mathrm{sech}^2(\sqrt{\lambda_{45}-(\lambda_1-\lambda_2+\lambda_3)}(x-x_0)),
\end{aligned}
$$
where $\lambda_1=0$ and
$$\lambda_2=\left\{
\begin{aligned}
&0, x<l\\
&U_0, x>l,
\end{aligned}
\right. \quad
\lambda_3=\left\{
\begin{aligned}
&0, x<0\\
&U_0, x>0.
\end{aligned}
\right. \quad
$$
The initial soliton velocity can be expressed by Riemann invariants $C=4\lambda_{45}+2(\lambda_1-\lambda_2+\lambda_3)$, and $C>2\lambda_1+4\lambda_3$, so the soliton will always tunnel as time develops.
Of course, it can also be explained by the strict hyperbolicity of KdV-Whitham system, and the trial soliton trajectory can be expressed by
\begin{equation*}
\left\{
\begin{aligned}
&\frac{d x_s}{dt}=v_{45}(0,\lambda_2(\frac{x_s(t)}{t}),\lambda_3(\frac{x_s(t)}{t}), \lambda_{45}),\\
&x_s^{initial}=x_0.
\end{aligned}
\right.
\end{equation*}
According to the transmission condition (\ref{transmission} ) and phase condition (\ref{invariance}), the amplitude of the trial soliton tunneling satisfies
$$
a_R=2(\overline{u}_M-\overline{u}_R)+a_M,\quad
\delta=x_R-x_M=(\sqrt{\frac{a_M}{a_R}}-1)x_M.
$$
\subsection{For $x_0>l$}
\begin{figure}
\centering
\includegraphics[width=11.0cm]{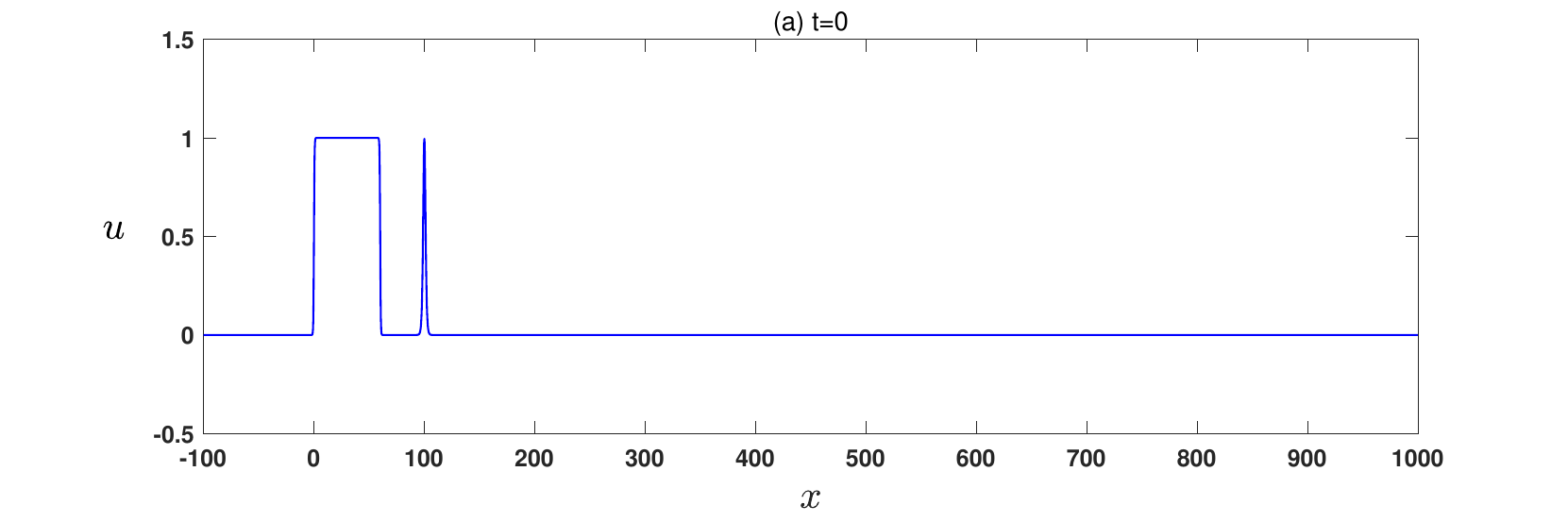}
\includegraphics[width=11.0cm]{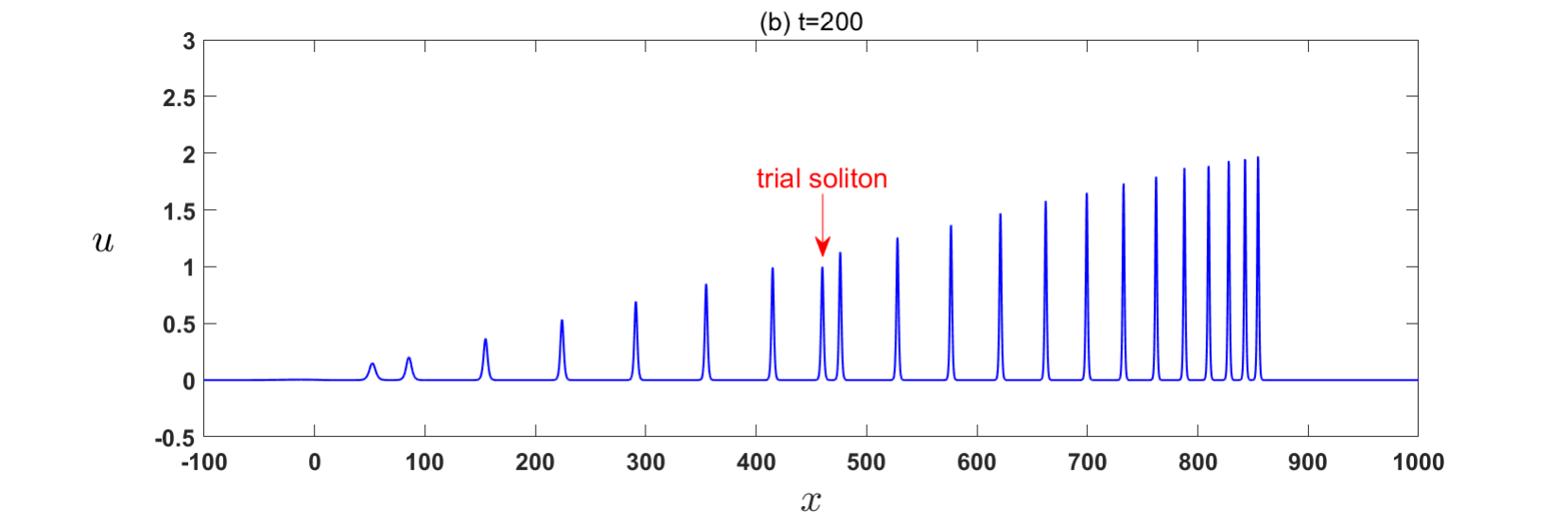}
\includegraphics[width=11.0cm]{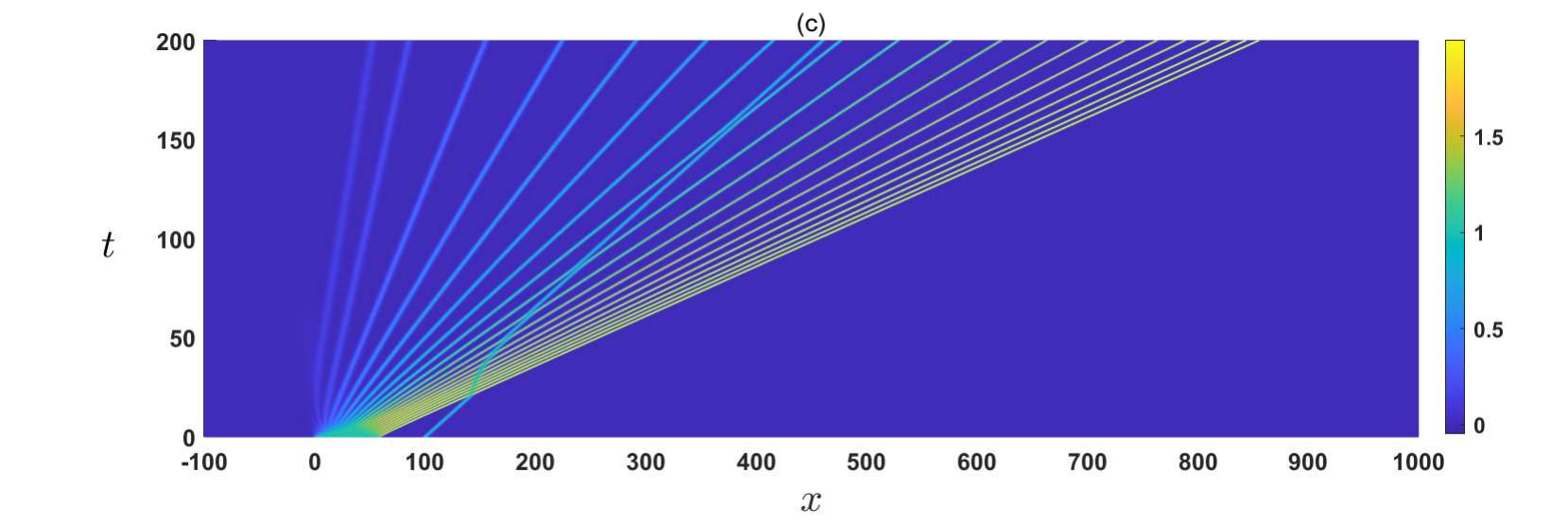}
\caption{{\protect\small The trial soliton is trapped in the interaction region when $x_0>l$. (a) Box type initial value and initial trial soliton; (b) The trial soliton is trapped completely; (d) The process of soliton trapping. The initial condition is set to be $U_0=1, l=20, x_0=60, a_L=1$.}}
\label{trapping-duibi-right-fig}
\end{figure}
Finally, consider the case that the location of initial trial soliton satisfies $x_0>l$, and the scale separation of the soliton and initial discontinuity requires $x_0\gg l$. If the speed of the initial soliton is very large, it will cause the soliton to have no interaction with the mean flow and just propagate to the right on the zero background. The critical condition is $a>2U_0$, which can be calculated by Whitham velocity. Obviously this is not what we are concerned about. In fact, since the traveling direction of soliton is always from left to right, and the left boundary of the entire wave region is $x=0$, this means that soliton tunneling will not occur.
\par
What is interesting here is the soliton trapping. Approximately initialize the box initial value and modulated soliton by using Riemann invariants in the form
$$
\begin{aligned}
u(x,0;x_0)=&\lambda_1-\lambda_4+\lambda_5\\
&+\left(2\lambda_{23}-2(\lambda_1-\lambda_4+\lambda_5)\right)\mathrm{sech}^2(\sqrt{\lambda_{23}-(\lambda_1-\lambda_4+\lambda_5)}(x-x_0)),
\end{aligned}
$$
where $\lambda_1=0$ and
$$\lambda_4=\left\{
\begin{aligned}
&0, x<l\\
&U_0, x>l,
\end{aligned}
\right. \quad
\lambda_5=\left\{
\begin{aligned}
&0, x<0\\
&U_0, x>0.
\end{aligned}
\right. \quad
$$
For the characteristic velocity $v_{23}=\lim_{\lambda_2\rightarrow \lambda_3}v_2^{(2)}=\lim_{\lambda_2\rightarrow \lambda_3}v_3^{(2)}$, one has
\begin{equation}
v_{23}=2(\lambda_1+\lambda_4+\lambda_5) -\frac{4(\lambda_5-\lambda_{23})\sqrt{\frac{(\lambda_4-\lambda_{23})(\lambda_{23}-\lambda_1)}{(\lambda_5-\lambda_1)(\lambda_5-\lambda_{23})}}}{\mathrm{Z}(\psi_{23},m_{23})},
\end{equation}
where $\mathrm{sin}\psi_{23}=\sqrt{\frac{(\lambda_{23}-\lambda_1)}{\lambda_4-\lambda_1}}$, and $m_{23}=\frac{\lambda_4-\lambda_1}{\lambda_5-\lambda_1}$.
We can calculate that the soliton is trapped in soliton train, co-propagating at a speed of $2\lambda_1+4\lambda_{23}$. The soliton trajectory can be obtained as follows
\begin{equation}
\left\{
\begin{aligned}
&\frac{d x_s}{dt}=v_{23}(0,\lambda_{23},\lambda_4(\frac{x_s(t)}{t}), \lambda_{5}(\frac{x_s(t)}{t})),\\
&x_s^{initial}=x_0.
\end{aligned}
\right.
\end{equation}
The numerical results are shown in Fig. \ref{trapping-duibi-right-fig}, where the interaction region generated by the square barrier appears as a soliton train with gradually increasing amplitude and propagating to the right. Those with high speed will elastically collide with the trial soliton until the trial soliton cannot be surpassed.

%

\section{Conclusions}
\label{sec:4}
This paper studies the soliton-mean flow interaction of the KdV equation with box type initial value plus a trial soliton. The box initial value is different from the single discontinuous step-like initial value, which directly leads to the interaction of the mean flow, that is, the region of soliton train. The emergence of the interaction region makes the results of the interaction between the trial soliton and the mean flow more abundant. This paper discusses various cases of trial soliton tunneling and trapping at different initial positions, and analyzes the tunneling and trapping conditions at different initial positions, see Table \ref{table}.
\begin{table}[htbp]
\footnotesize
\caption{Conditions for soliton tunneling and soliton trapping}\label{table}
\begin{center}
  \begin{tabular}{|c|c|c|c|} \hline
  \bf Initial position   & \bf Soliton tunneling & \bf Soliton trapping & \bf No interaction \\ \hline
    $x_0<0$ & $a>2U_0$ & $a<2U_0$ &/ \\ \hline
    $0<x_0<l$ & always & / &/ \\ \hline
    $x_0>l$ & / & $a<2U_0$ & $a>2U_0$\\ \hline
  \end{tabular}
\end{center}
\end{table}
\par
Within the framework of Whitham modulation theory, both the transmission and phase conditions for the interaction of solitons with the mean flow are derived, and a comparison between the prediction results and the numerical results is given in detail, in which the error is controlled within an acceptable range. In addition to constructing Riemann invariants to describe the interaction between soliton and mean flow, the soliton tunneling and soliton trapping are also explained from the behaviors of the spectrums.

\section*{Acknowledgments}
Support is acknowledged from the National Natural Science Foundation of China, Grant No. 12371247.	
	
	\bibliographystyle{amsplain}

\begin{thebibliography}{10}
		
\bibitem{Zabusky-1965} N. J. Zabusky and M. D. Kruskal, Interactions of ``solitons" in a collisionless plasma and the recurrence of initial states, Phys. Rev. Lett., 15 (1965), pp. 240-243.
\bibitem{Lax-1968} P. D. Lax, Integrals of nonlinear equations of evolution and solitary waves, Commun. Pure Appl. Math., 21 (1968), pp. 467-490.
\bibitem{Gardner-1967}C. S. Gardner, J. M. Gardner, M. D. Kruskal and R. M. Miura,  Method for solving the Korteweg-de Vries equation, Phys. Rev. Lett., 19 (1967), pp. 1095-1097.
\bibitem{Miura-1976}R. Miura, The Korteweg-de Vries equation: a survey of results, SIAM review,  18 (1976), pp. 412-459.
\bibitem{Whitham-1965}G. B. Whitham,  Non-linear dispersive waves, Proc. R. Soc. Lond. Ser. A, 283 (1965), 238.
\bibitem{AMK-1997}A. M. Kamchatnov, New approach to periodic solutions of integrable equations and nonlinear theory of modulational instability, Phys. Rep. 286 (1997), pp. 199-270.
\bibitem{Kodama-1999}Y. Kodama, The Whitham equations for optical communications: Mathematical theory of NRZ, SIAM J. Appl. Math., 59 (1999), pp. 2162-2192.
\bibitem{Biondini-2006}G. Biondini and Y. Kodama, On the Whitham equations for the defocusing nonlinear Schrodinger equation with step initial data, J. Nonlinear Science, 16 (2006), pp. 435-481.
\bibitem{Hoefer-2007}M. A. Hoefer and M. J. Ablowitz, Interactions of dispersive shock waves, Physica D, 236 (2007), pp. 44-64.
\bibitem{Pavlov-2022}E. V. Ferapontov and M. V. Pavlov, Kinetic equation for soliton gas: integrable reductions, J. Nonlinear Science, 32 (2022), 26.
\bibitem{Wang-PD-2022}R. Gong and D. S. Wang, Formation of the undular bores in shallow water generalized Kaup-Boussinesq model, Physica D, 439 (2022), 133398.
\bibitem{Wang-Stud-2022}Y. Liu and D. S. Wang, Exotic wave patterns in Riemann problem of the high-order Jaulent-Miodek equation: Whitham modulation theory, Stud. Appl. Math. 149 (2022), pp. 588-630.
\bibitem{Oliver-2009}O. B\"{u}hler, Waves and Mean Flows, Cambridge Monographs on Mechanics, Cambridge University Press, 2009.
\bibitem{Gurevich-1974}A. V. Gurevich and L. P. Pitaevskii, Nonstationary structure of a collisionless shock wave, Soviet J. Experimental and Theoretical Physics, 38 (1974), pp. 291-297.
\bibitem{Luo-stud-2023} M. J. Ablowitz,  J. T. Cole,  G. A. El,  M. A. Hoefer and X. Luo, Soliton-mean field interaction in Korteweg-de Vries dispersive hydrodynamics, Stud. Appl. Math., 151 (2023),  pp. 795-856.
\bibitem{Hirschfelder-1974}J. O. Hirschfelder, A. C. Christoph and W. E. Palke, Quantum mechanical streamlines. I. Square potential barrier, Journal of Chemical Physics,  61 (1974), pp. 5435-5455.
\bibitem{Jenkins-2014}R. Jenkins and K. D. T-R. McLaughlin, Semiclassical limit of focusing NLS for a family of square barrier initial data, Comm. Pure Appl. Math., 67 (2014), pp. 246-320.
\bibitem{EL-PRL-2018} M. D. Maiden, D. V. Anderson, N. A. Franco, G. A. El and M. A. Hoefer, Solitonic Dispersive Hydrodynamics: Theory and Observation, Phys. Rev. Lett., 120 (2018), 144101.
\bibitem{EL-PRE-2018} P. Sprenger, M. A. Hoefer and G. A. El, Hydrodynamic optical soliton tunneling, Phys. Rev. E, 97 (2018), 032218.
\bibitem{EL-JFM-2021}K. Van der Sande, G. A. El, and M. A. Hoefer, Dynamic soliton-mean flow interaction with non-convex flux, J. Fluid Mech., 928 (2021), A21.
\bibitem{EL-Chaos-2005}G. A. El, Resolution of a shock in hyperbolic systems modified by weak dispersion, Chaos, 15 (2005), 037103.
\bibitem{FFM-CPAM-1980}H. Flaschka, M. G. Forest and D. W. McLaughlin, Multiphase averaging and the inverse spectral solution of the Korteweg-de Vries equation, Comm. Pure Appl. Math., 33 (1980), pp. 739-784.
\bibitem{EL-Chaos-2002}G. A. El, and R. H. J. Grimshaw, Generation of undular bores in the shelves of slowly-varying solitary waves, Chaos,  12 (2002) pp. 1015-1026.
\bibitem{Karpman-1967}V.I. Karpman, An asymptotic solution of the Korteweg-de Vries equation,  Phys. Lett. A, 25 (1967), pp. 708-709.
\bibitem{Tricomi-1961}F. G. Tricomi, Differential equations, Courier Corporation, 2012.



		
		
		
		
		
		
		
		
		
		
	\end{thebibliography}

\end{document}